\documentclass[11pt,prd,letterpaper,aps]{article}
\pdfoutput=1

\usepackage{fancyhdr}
\usepackage{relsize}
\usepackage{amsmath,amssymb}
\usepackage{color}
\usepackage{epsfig}
\usepackage{graphicx}               
\usepackage{url,cite}
\usepackage{color}
   \usepackage{hyperref}
   \definecolor{db}{RGB}{0,0,221}

\usepackage[T1]{fontenc}
\usepackage{textcomp}
\usepackage{pifont}

\usepackage{caption}
\newcommand{\nc}{\newcommand}

\nc{\beq}{\begin{equation}}
\nc{\eeq}{\end{equation}}
\nc{\beqa}{\begin{eqnarray}}
\nc{\eeqa}{\end{eqnarray}}
\nc{\bea}{\begin{eqnarray}}
\nc{\eea}{\end{eqnarray}}
\nc{\ra}{\rightarrow}
\nc{\lsim}{\begin{array}{c}\,\sim\vspace{-21pt}\\< \end{array}}
\nc{\gsim}{\begin{array}{c}\sim\vspace{-21pt}\\> \end{array}}
\nc{\Tr}{{\rm Tr}}
\nc{\slsh}{\slash\hspace*{-0.22cm}}

\def\be{\begin{equation}}
\def\ee{\end{equation}}
\def\bea{\begin{eqnarray}}
\def\eea{\end{eqnarray}}
\def\bit{\begin{itemize}}
\def\eit{\end{itemize}}

\def\to{\rightarrow}

\definecolor{red}{rgb}{0.75,0,0}

\title{
\vspace*{-2.3cm}
\begin{flushright}
\normalsize{
  }
\end{flushright}
\vspace{1.5cm}
\Large
\textbf{
Dark Matter Searches with a Mono-$Z^\prime$ Jet
}
\vspace*{1.0cm}
}

\author{\normalsize \bf Yang Bai,$^{a}$ James Bourbeau,$^{a}$ and Tongyan Lin$^{b}$
\vspace{5mm}
\\
$^{a}$ \normalsize\emph{Department of Physics, University of Wisconsin, Madison, WI 53706, USA}  \vspace{1mm} \\
$^{b}$ \normalsize\emph{Kavli Institute for Cosmological Physics, University of Chicago, Chicago, IL 60637, USA}
}

\date{}

\begin{document}
\setcounter{page}{0}
\maketitle

\vspace*{1cm}
\begin{abstract}
 \normalsize{
 We study collider signatures of a class of dark matter models with a GeV-scale dark $Z^\prime$. At hadron colliders, the production of dark matter particles naturally leads to associated production of the $Z^\prime$, which can appear as a narrow jet after it decays hadronically. Contrary to the usual mono-jet signal from initial state radiation, the final state radiation of dark matter can generate the signature of a mono-$Z^\prime$ jet plus missing transverse energy. Performing a jet-substructure analysis to tag the $Z^\prime$ jet, we show that these $Z^\prime$ jets can be distinguished from QCD jets at high significance. Compared to mono-jets, a dedicated search for mono-$Z^\prime$ jet events can lead to over an order of magnitude stronger bounds on the interpreted dark matter-nucleon scattering cross sections.
 }
\end{abstract}

\thispagestyle{empty}
\newpage

\setcounter{page}{1}

\baselineskip16pt

\vspace*{-0.5cm}
\section{Introduction}
\label{sec:intro}
After the discovery of the Higgs boson in 2012, searching for physics beyond the Standard Model (SM) has become the highest priority at the Large Hadron Collider (LHC). One of the most important new BSM particles is dark matter, whose existence has long been established from astrophysical observations. In spite of a long history of searching for dark matter particles from direct detection, indirect detection and accelerator-based experiments, there is still no clear evidence for the particle nature of dark matter. 

The accelerator-based searches for dark matter can be traced back to the Super Proton Synchrotron (SPS) at CERN. The UA1 collaboration reported evidence for events with mono-jet and mono-photon plus large missing transverse energy~\cite{Arnison:1984qu}. This triggered an interesting dark matter interpretation in the supersymmetric framework~\cite{Ellis:1984jc}, although it was later explained away by the production of weak gauge bosons plus additional jets~\cite{Ellis:1985ig}. Since then, collider searches for dark matter have become standard at LEP, Tevatron and the LHC. 

In many searches for dark matter at colliders, the dark matter particle is embedded into the framework of supersymmetry and the corresponding signatures are model-dependent. For instance, pair-production of two squarks can lead to a final state of  two jets plus missing transverse energy, $E_T^{\rm miss}$, after each squark decays into a quark and a neutralino (the dark matter candidate particle). Signatures like multi-jets plus $E_T^{\rm miss}$, multi-leptons plus $E_T^{\rm miss}$, $t\bar{t}+E_T^{\rm miss}$ and $b\bar{b}+E_T^{\rm miss}$ have dedicated searches at both the CMS and ATLAS collaborations at the 8 TeV LHC. Moving from the LHC Run I to the LHC Run II, those signatures can teach us about the supersymmetric spectrum as well as how dark matter interacts with SM particles~\cite{Cahill-Rowley:2014twa}. 

A less model-dependent approach is to consider effective higher-dimensional operators or simplified models to describe dark matter interactions. For the effective operator analysis, a large class of signatures have been proposed and searched for at the LHC: mono-jet~\cite{Goodman:2010ku,Bai:2010hh}, mono-photon~\cite{Birkedal:2004xn,Gershtein:2008bf,Fox:2011fx}, mono-$W$~\cite{Bai:2012xg,Crivellin:2015wva}, mono-$Z$~\cite{Petriello:2008pu,Carpenter:2012rg,Yu:2014ula,Crivellin:2015wva}, mono-Higgs~\cite{Petrov:2013nia,Carpenter:2013xra,Berlin:2014cfa} and mono-$b$~\cite{Lin:2013sca}. For simplified dark matter models~\cite{Abdallah:2014hon}, one has a jets plus missing energy signature in the quark-portal dark matter models~\cite{Chang:2013oia,Bai:2013iqa,DiFranzo:2013vra,Batell:2013zwa,Papucci:2014iwa,Garny:2014waa,Gomez:2014lva,Haisch:2015ioa}, a two leptons plus missing energy signature in the lepton-portal dark matter models~\cite{Bai:2014osa,Chang:2014tea,Agrawal:2014ufa,Bell:2014tta,Yu:2014mfa}, or just visible dilepton signatures if dark matter interacts with both quarks and leptons~\cite{Altmannshofer:2014cla}.  The simplified model of dark matter coupled to a new $Z'$ has also been studied extensively~\cite{An:2012va,An:2012ue,Frandsen:2012rk,Arcadi:2013qia,Alves:2013tqa,Busoni:2014sya,Alves:2015pea}.

\begin{figure}[t!]
\begin{center}
\includegraphics[width=0.5\textwidth]{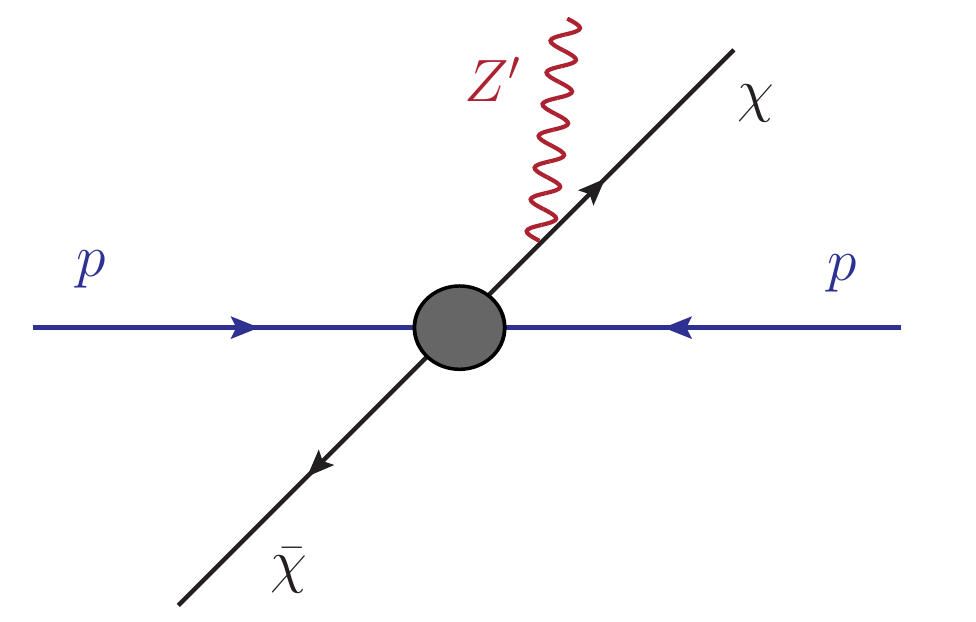}
\caption{An illustrative Feynman diagram for the mono-$Z^\prime$ signature at hadron colliders. The $Z^\prime$ is mainly produced from dark matter final state radiation. For a GeV-scale $Z^\prime$ decaying to hadrons, this gives a unique mono-$Z^\prime$ jet signature. }
\label{fig:zprime-FSR}
\end{center}
\end{figure}

While many existing mono-$X$ studies have concentrated on identifying signals using the initial state radiation (ISR) of partons inside an accelerated proton, less attention has been paid to potential dark matter final state radiation (FSR). The basic process is that dark matter is pair produced, after which one of the particles can radiate  a dark $Z^\prime$; a cartoon of this is illustrated in Fig.~\ref{fig:zprime-FSR}. The $Z^\prime$ from FSR can decay back to SM particles and behave as a visible object in the collider, while there can still be substantial missing transverse momentum from the dark matter particles. 

Here we focus on the possibility that the dominant decay of the $Z^\prime$ is into quarks. When the $Z^\prime$ mass scale is light (GeV-scale), then there are two important effects we identify. First, the hadronic decay of a boosted $Z^\prime$ gives a new collider signature: the $Z'$ appears as a jet with a very narrow cone of radiation and a small multiplicity of charged particles. We refer to these as $Z'$-jets, and show that these can be distinguished at high significance from QCD jets. Second, the rate for dark matter FSR of $Z^\prime$ jets can be larger than the rate for ISR jets. Taking advantage of both effects, we demonstrate that a dedicated collider analysis based on the mono-$Z^\prime$ signature will dramatically improve our understanding of the dark matter interactions with visible particles.  

In this paper, we categorize collider signatures with dark matter radiating its own force carrier, for simplicity assumed to be a spin-one vector boson. We concentrate on an Abelian dark matter sector, with a GeV-scale $Z^\prime$. Due to the kinematic constraints, the $Z^\prime$ will decay into only a few hadrons. For the examples in our paper, the $Z^\prime$ will mainly decay into two or three mesons, of which two are charged. By requiring large missing transverse momentum, the $Z^\prime$ particle is boosted and the  decay products are highly collimated. This mono-$Z^\prime$ jet can be differentiated from a QCD jet using a jet substructure analysis. 

Complementary work on radiation of heavier $Z'$s and different decay channels can be found in Refs.~\cite{Autran:2015mfa,Gupta:2015lfa}. For heavier $Z'$s decaying hadronically, one can search for missing transverse momentum plus a resonance in the invariant mass of the two jets to reduce the SM backgrounds. The dilepton resonance plus missing transverse momentum signature probes $Z'$s that decay leptonically.
We also note that a non-Abelian GeV-scale dark sector can naturally result in a cascade of gauge bosons. The latter case has been studied in the context of lepton jets~\cite{ArkaniHamed:2008qp,Bai:2009it,Cheung:2009su,Baumgart:2009tn,Aad:2014yea} as well as jets with hadronic shower products that nevertheless could be distinguished from QCD jets~\cite{Bai:2013xga,Schwaller:2015gea,Cohen:2015toa}.  

Our paper is organized as follows. We begin in Section~\ref{sec:jet-sub} with a general discussion of the collider signature of a GeV-scale $Z'$ jet, employing jet substructure variables for $Z'$-jet tagging. In Section~\ref{sec:secluded} we discuss secluded dark $Z^\prime$ models. We consider the elastic dark matter case in Section~\ref{sec:elastic} and the inelastic dark matter case in  Section~\ref{sec:idm}, and compute the sensitivity at the 14 TeV LHC. In Section~\ref{sec:public} we turn to a ``public'' dark $Z^\prime$ model, where SM fermions are also charged under the $Z'$, and determine the projected LHC sensitivity. Finally, we conclude our paper in  Section~\ref{sec:conclusion}.  

\section{Mono-$Z^\prime$ Jets}
\label{sec:jet-sub}
Light $Z^\prime$s decaying hadronically give rise to mono-$Z^\prime$ jets, which have different characteristics compared to an ordinary QCD jet. Since the $Z^\prime$ is produced in association with large $E_T^{\rm miss}$, it is highly boosted, leading to a jet with a small-radius cone of radiation, of typical size $\sim M_{Z^\prime}/p_T(Z^\prime)$. Furthermore, the dominant $Z'$ decay leads to two tracks and favors a smaller jet mass. Meanwhile, high $p_T$ QCD jets on average have a larger track multiplicity and a larger jet mass. This motivates us to employ jet-substructure techniques to distinguish the mono-$Z^\prime$ jet from an ordinary QCD jet. 

Our mono-$Z^\prime$ jet has many similarities to the hadronic $\tau$ (or $\tau_h$) object: both have only a few hadrons confined in a small geometric region and have low invariant mass. In our implementation of light $Z'$ tagging, we therefore adopt a number of the substructure variables used in $\tau$-tagging at the LHC~\cite{ATLAS-CONF-2013-064,Aad:2014rga}. Due to charge conservation and the low mass of the $Z'$, our mono-$Z^\prime$ jet prefers to have two charged particles in the final state, so it should behave dominantly as a {\it two-prong} object instead of one-prong or three-prong structure like $\tau_h$. Depending on the ability to resolve the number of tracks, the $\tau_h$ can comprise part of the background for the mono-$Z^\prime$ signal.

\begin{figure*}[tb!]
\begin{center}
 \includegraphics[width=0.47\textwidth]{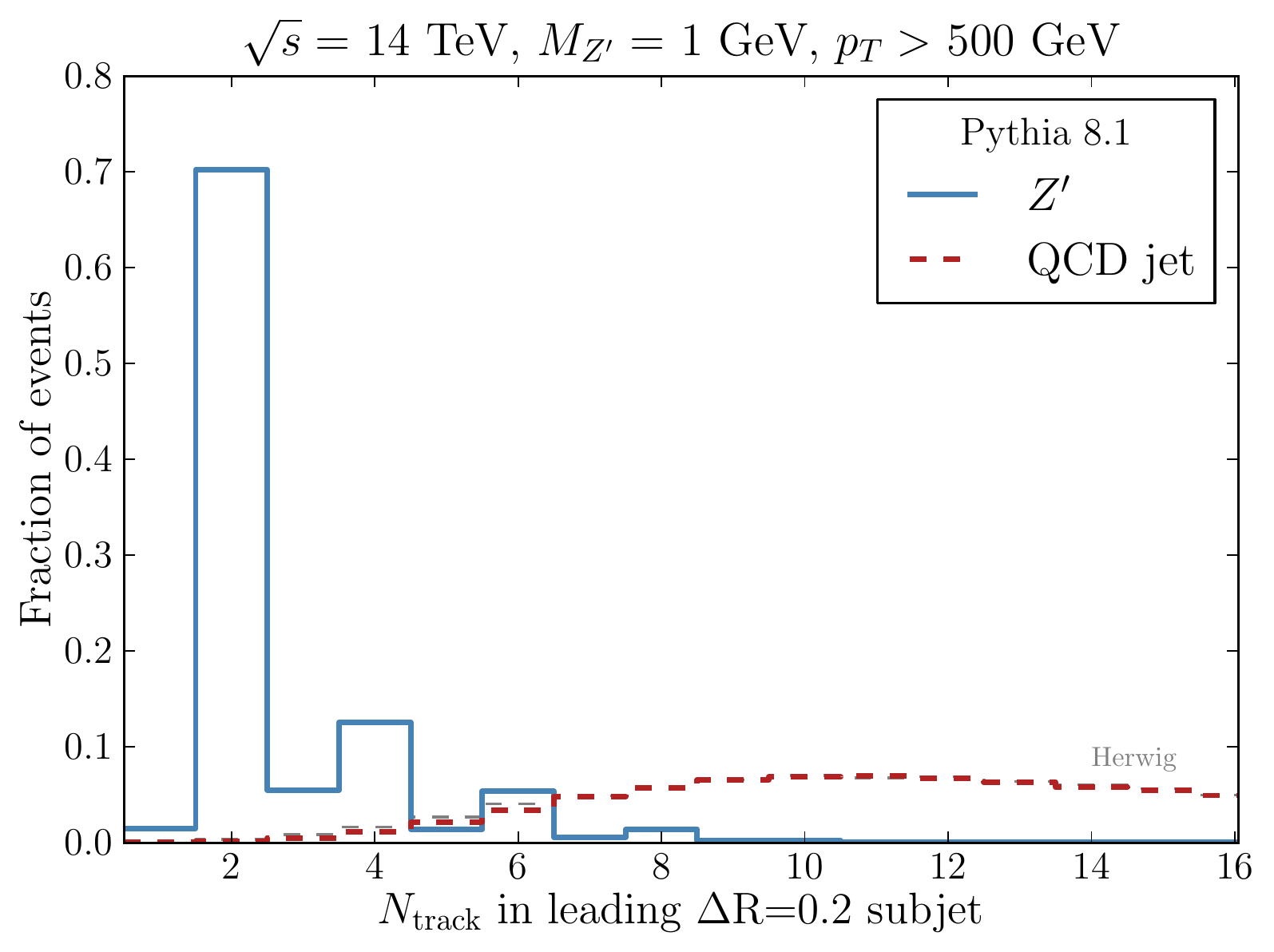}
 \includegraphics[width=0.47\textwidth]{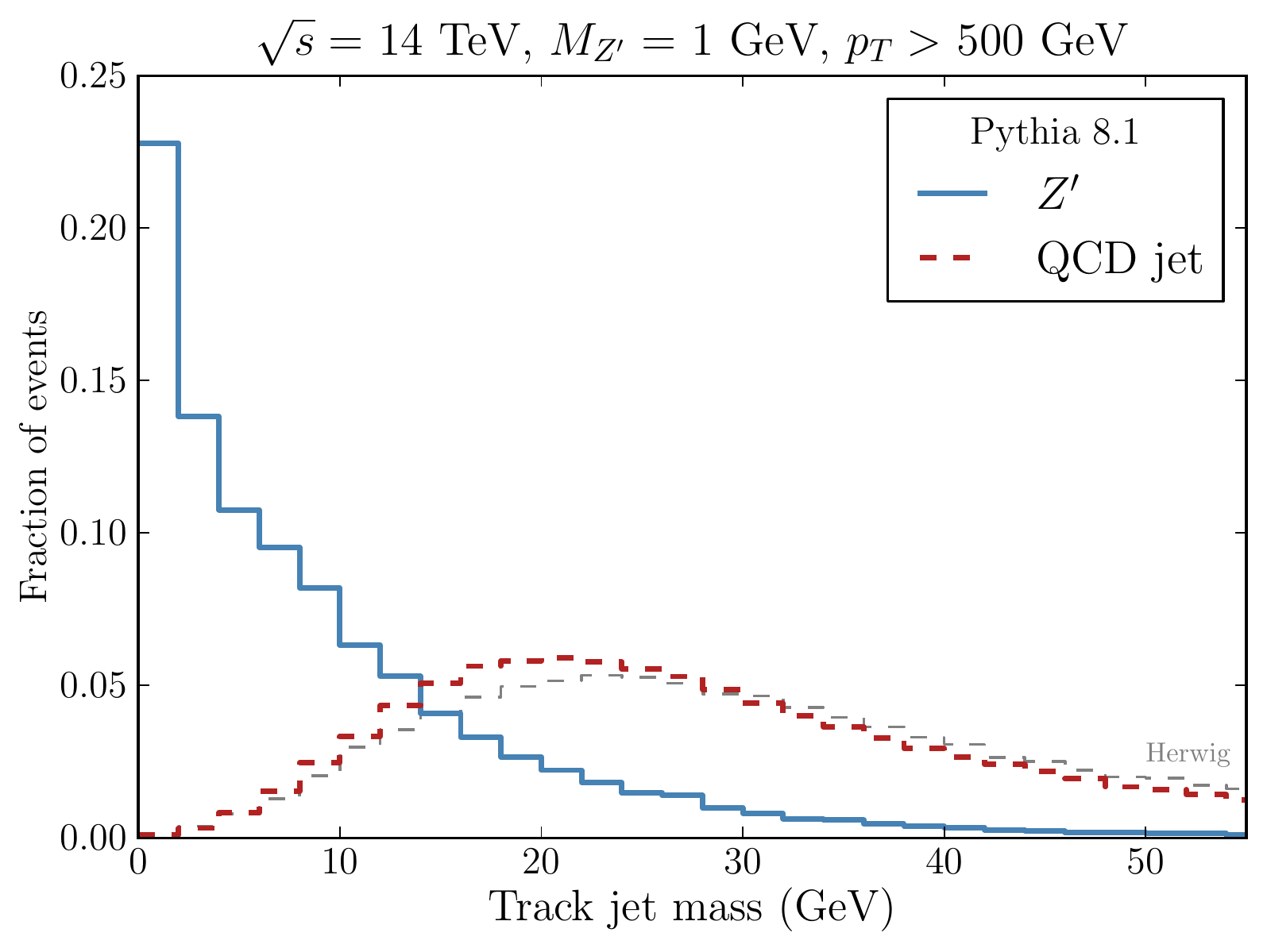}
 \includegraphics[width=0.47\textwidth]{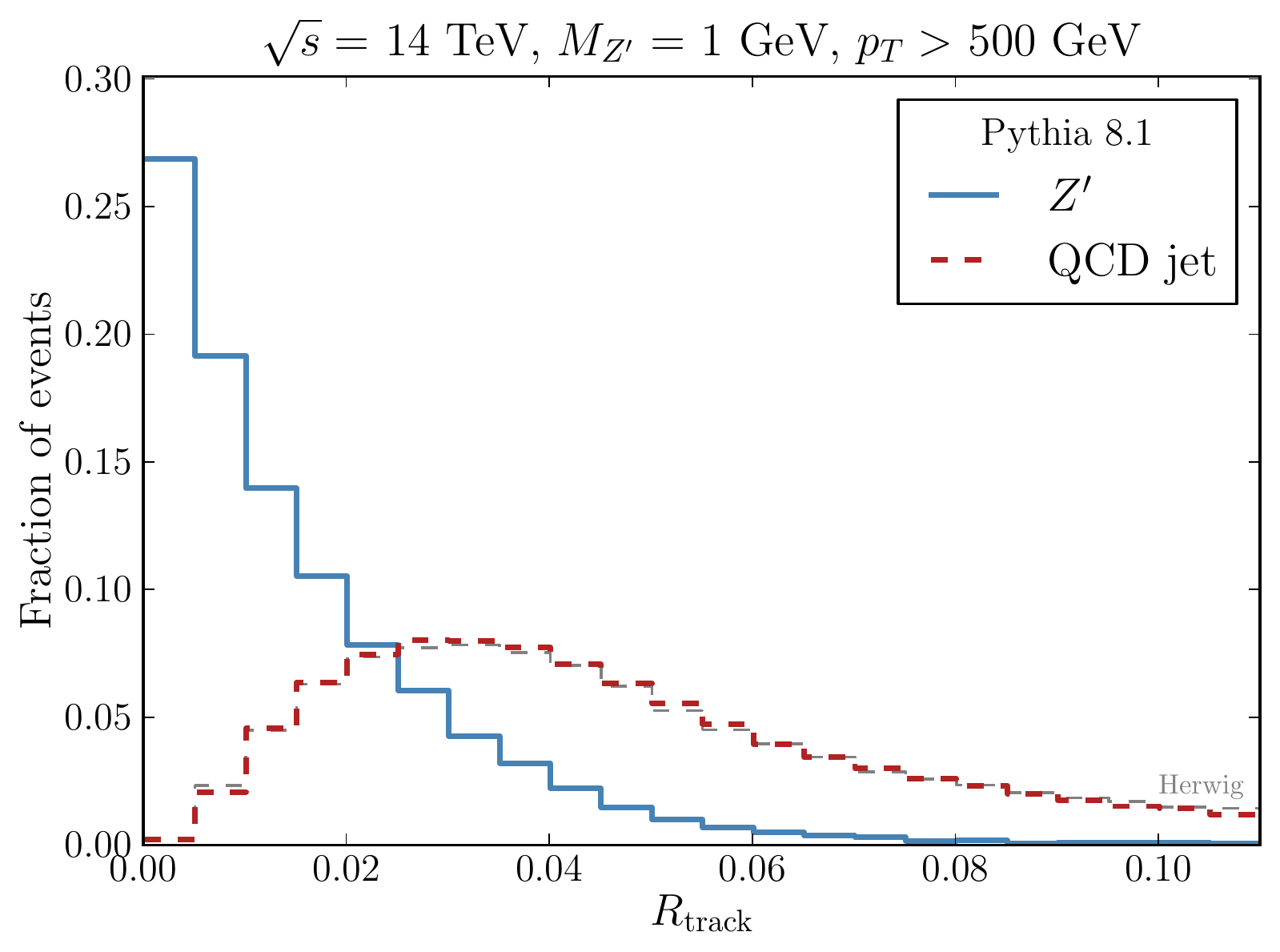}
 \includegraphics[width=0.47\textwidth]{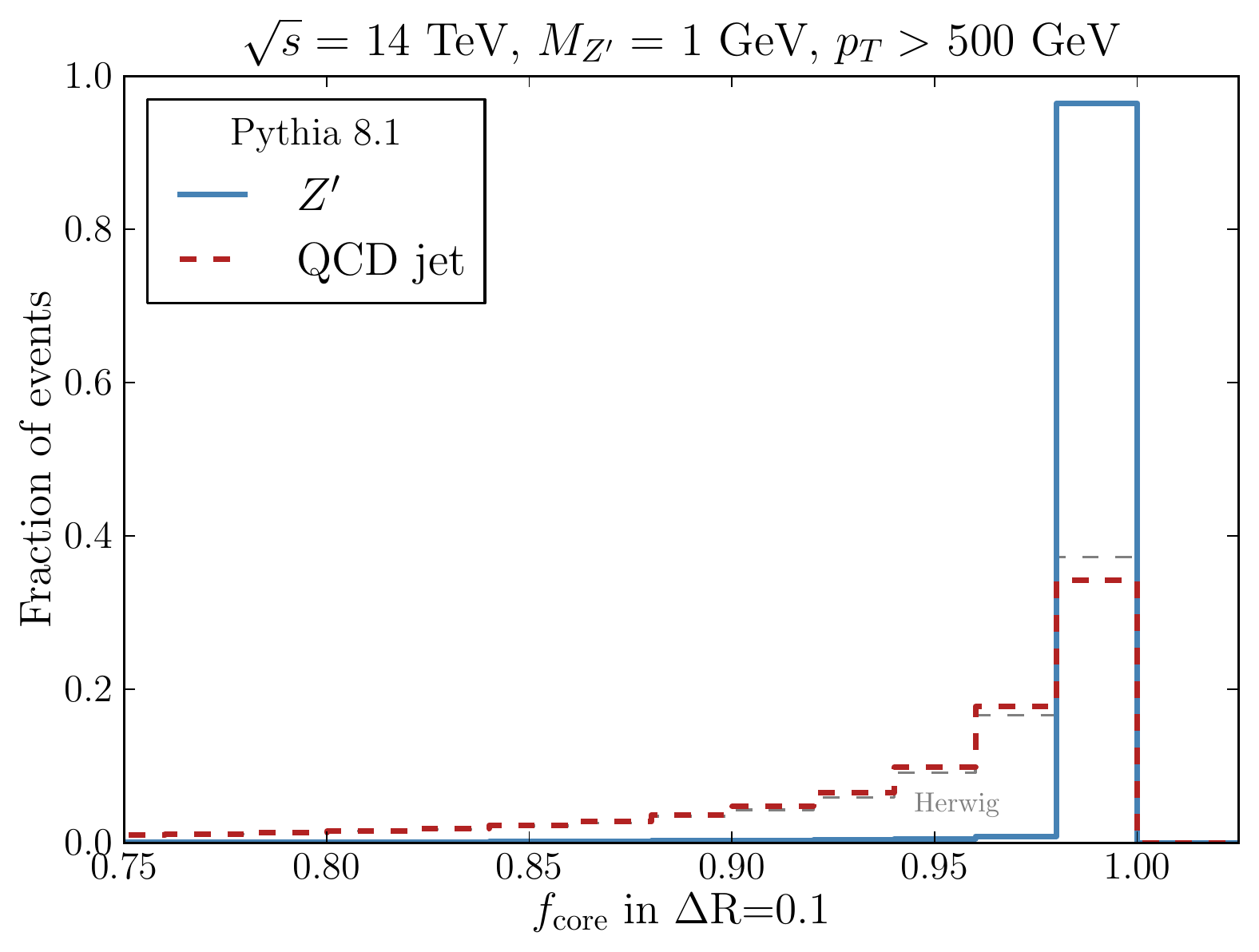}
\end{center}
\caption{  Jet substructure variables with close analog in $\tau$-tagging, for the 14 TeV LHC. The $Z'$ jet shown here is assuming an isospin-violating vector-coupling of the $Z'$ to light quarks. We show results using \texttt{Pythia 8.1} for both $Z'$ jets and QCD jets, and also using \texttt{Herwig++} for QCD jets. \label{fig:jetsub}}
\end{figure*}

Motivated by the $\tau$-tagging variables in Ref.~\cite{ATLAS-CONF-2013-064}, we consider the following four primary variables for mono-$Z^\prime$ jet tagging. Here our reconstruction algorithm is seeded from jet-objects reconstructed using the anti-$k_t$ algorithm~\cite{Cacciari:2008gp} with $R=0.4$. 
\begin{itemize}
	\item $N_\textrm{track}$, the number of tracks in the leading $\Delta R=0.2$ subjet. 
		\item $m_\textrm{track}$, the track jet mass. 
	\item Track radius $R_\textrm{track}$, the $p_T$-weighted track width:
	\begin{equation}
		R_\textrm{track} = \frac{  ‎‎‎‎\sum_{i, {\rm tracks}}^{\Delta R_i \leq 0.4}  ~  p_{T,i}\,\Delta R_i }{ \sum_{i,{\rm tracks}}^{\Delta R_i \leq 0.4} ~ p_{T,i} } \,.
	\end{equation}	
	\item $f_\textrm{core}$, which parametrizes the $p_T$-fraction of the leading subjet:
	\begin{equation}
		f_\textrm{core} \equiv \frac{ \sum^{\Delta R_i < 0.1}_{i} ~ p_T^i }{ \sum^{\Delta R_i < 0.2}_i ~ p_T^i} \,.
	\end{equation}
\end{itemize} 
Note that for $\tau$-tagging the definition of $\Delta R_i$ is relative to the $\tau$ intermediate axis, which is the axis defined by the inner $\Delta R < 0.2$ of the clustered jet. For simplicity, we define $\Delta R_i$ relative to the jet axis.

We have focused on track-based variables here in order to avoid the additional complications of pileup and calorimeter energy resolution. Additional variables may be able to further improve background rejection, but would be subject to these systematic uncertainties: for example, defining a scaled track jet mass $m_{\rm track} \times p_T/p_T^{\rm tracks}$ or including an additional observable characterizing the fraction of jet $p_T$ in charged tracks helps distinguish between $Z'$ and QCD jets, and may additionally improve rejection of QCD jets by up to a factor of 2. 

For $Z'$ jets with $p_T > 500$ GeV and $M_{Z^\prime}=1$~GeV, the separation of the tracks is on the order of $10^{-3}$ which is similar to the angular resolution of the ATLAS inner tracker~\cite{Aad:2008zzm,Aad:2010bx}. Although this presents an experimental challenge, the variables considered here are primarily sensitive to the distribution of the radiation and can be applicable even if individual tracks are difficult to resolve.

Distributions of the observables above for $Z'$-jets and QCD jets are shown in Fig.~\ref{fig:jetsub}, where there is a clear difference in the properties of the two objects. (For $\tau_h$, the distributions are very similar to that of $Z'$-jets, except that events primarily have $N_{\rm track} = 1$ or $N_{\rm track} = 3$.)
To account for the experimental resolution of tracks, we have applied a simple smearing of track $p_T$ with $\delta p_T/p_T = 0.05$~\cite{Aad:2010bx}. This has the largest effect on the track jet mass and the track radius, while barely affecting $f_\textrm{core}$. Note that since the primary distinguishing observables are track-based, and since the observables are highly correlated, we neglect the effects of pileup and calorimeter energy resolution in $f_\textrm{core}$.

\begin{table*} [t!]
 \renewcommand{\arraystretch}{1.25}
\centering
  \begin{tabular}{ | c | c | c | c ||  c | c | }
    \hline
	   $f_\textrm{core} \ge$  & $m_\textrm{track} \le $  &  $R_\textrm{track}\le$  &  $N_\textrm{track}\le$  &  Signal Eff. & Background Eff. \\ \hline \hline
	  0.95   & 10 GeV          &      0.01        &   4     &  0.36   &   0.001 \\  \hline
	 0.9   & 20 GeV          &      0.01        &   4     &  0.40     &    0.002 \\ \hline
	 0.9   & 20 GeV          &      0.02        &   2     &  0.48     & 0.001    \\  \hline
	 0.9  &  20 GeV          &      0.02        &   4     &  0.60     & 0.005    \\  \hline
	 0.9   & 20 GeV          &      0.02        &   6     &  0.63     & 0.019    \\  \hline
  \end{tabular}
  \caption{Mono-$Z'$ tagging efficiencies for benchmark values of the cuts, with $p_T\ge$ 500 GeV at the 14 TeV LHC. The signal is for $M_{Z'} = 1$ GeV, while the background is for QCD jets. For the cuts in the last two rows, there is only a small difference between using \texttt{Pythia} or \texttt{Herwig++}.}
  \label{table:eff}
\end{table*}

For the results in Fig.~\ref{fig:jetsub} and as the default in this section, we use \texttt{Pythia}~\cite{Sjostrand:2007gs} to shower and hadronize the parton-level events, including the $Z'$ decay. Strictly speaking, the hadronization model used  in \texttt{Pythia}~\cite{Sjostrand:2007gs} may not be valid at scales of 1 GeV, while our chiral perturbation theory analysis (discussed in the following section) is only accurate at scales well below 1 GeV. However, as the results show, the behavior of the jet substructure observables agrees with our intuition based purely on the kinematics of the event and charge conservation, and taking into account the effects of  the $p_T$ resolution of the detector and contamination from additional soft  radiation in each event. We have also compared results using \texttt{Herwig++}~\cite{Bahr:2008pv}, which employs a different hadronization model. The jet-substructure variables for QCD jets are very similar between the two models, as shown in Fig.~\ref{fig:jetsub}. For $Z^\prime$ jets, we find that \texttt{Pythia} appears to give a better match onto the expectation from chiral perturbation theory for $N_{\rm track}$ and so we primarily use \texttt{Pythia}.

To demonstrate the feasibility of light $Z'$ tagging, we consider cuts in the four observables above  in Table~\ref{table:eff} and show that efficiencies comparable to that in $\tau$-tagging are possible. While extremely high background rejection ($\sim 10^3$) may be achievable with stringent cuts on $m_\textrm{track}$ and $R_\textrm{track}$, the distributions of these observables at very low values are more sensitive to the specifics of $Z'$ branching ratios and hadronization model. As a result, we consider relatively conservative cuts on the observables in the last two rows of Table~\ref{table:eff}, in order to reduce the uncertainty associated with this model-dependence.  We then find that the most important variables that allow a robust rejection of QCD jets at the percent level or better are $N_\textrm{track}$, followed by  $R_\textrm{track}$.

\begin{figure*}[tb!]
\begin{center}
 \includegraphics[width=0.47\textwidth]{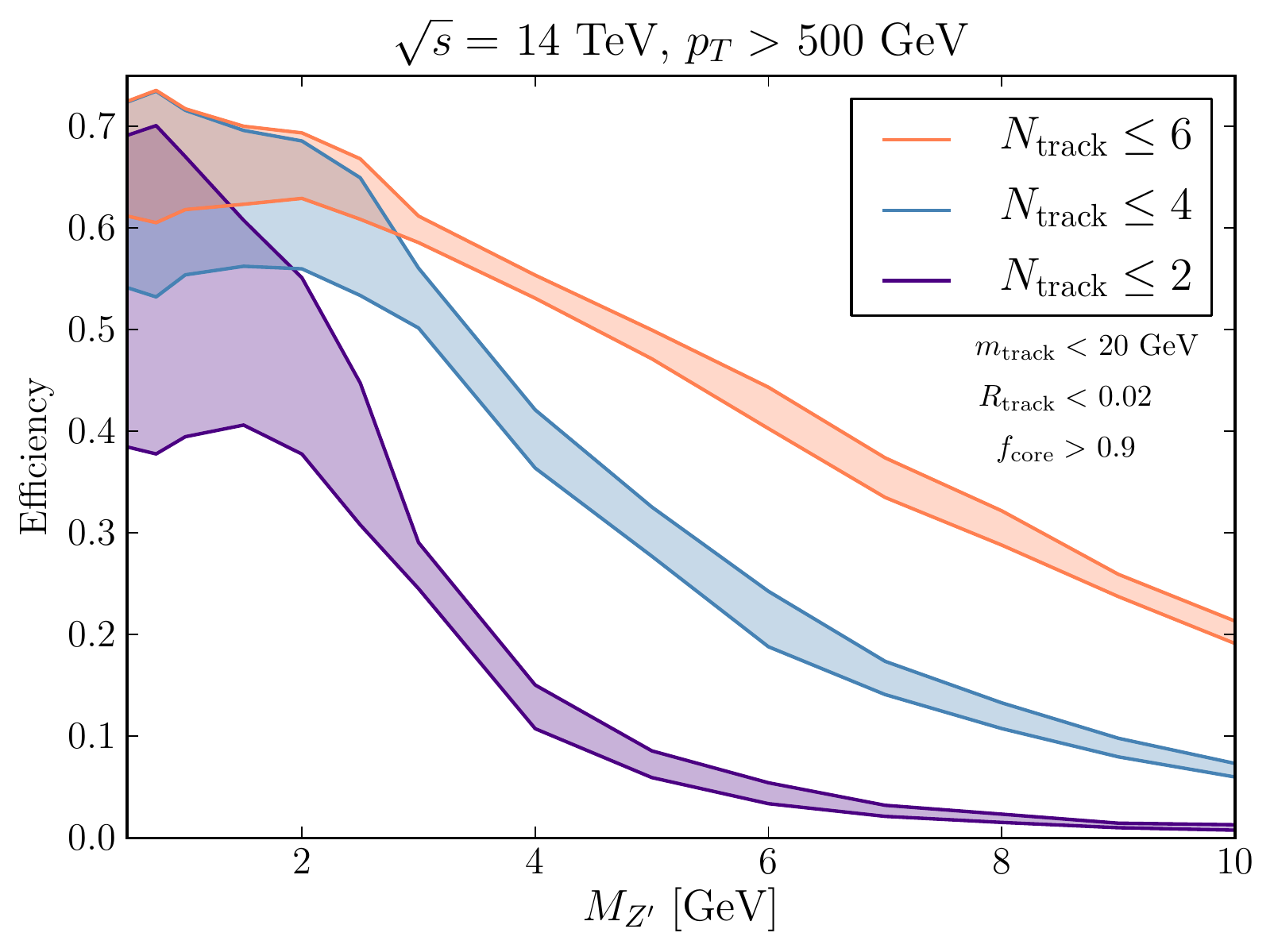}
 \includegraphics[width=0.47\textwidth]{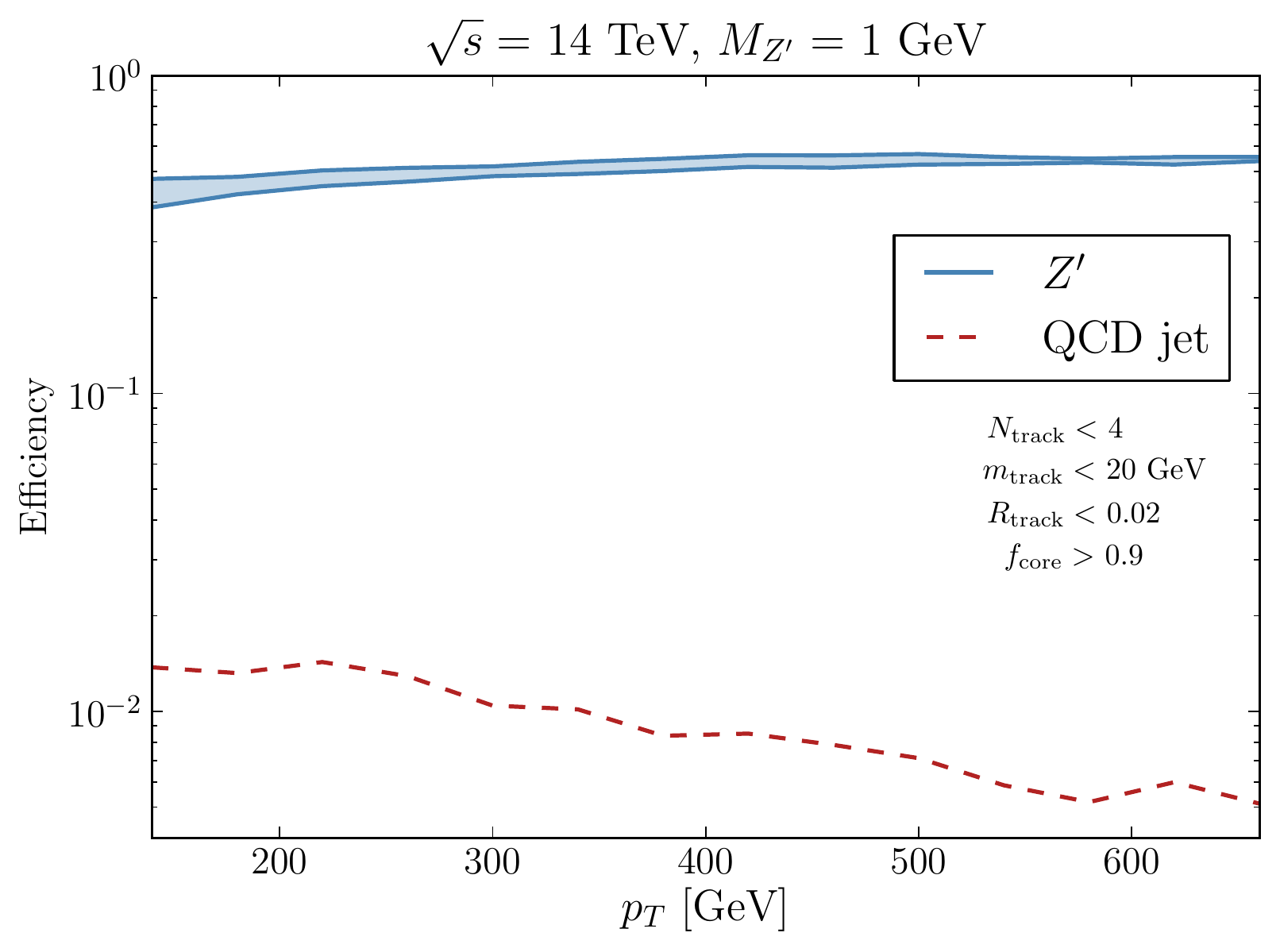}
\end{center}
\caption{ Left panel: the dependence of the $Z'$ tagging efficiency on the $Z'$ mass, for various $N_{\rm{track}}$ cuts.  Right panel: the tagging efficiency as a function of the jet $p_T$ for both $Z'$ and QCD jets. In both figures, the shaded bands show the range of results obtained by modifying the structure of the $Z'$ coupling to light quarks,  as described in the text.
 \label{fig:jetsub2}}
\end{figure*}

Furthermore, to distinguish mono-$Z^\prime$ jets from QCD jets, we only require $N_{\rm track} \leq 4$. For real $\tau_h$, application of the cuts in Table~\ref{table:eff} then leads to a similar efficiency as for our signal. In order to reduce the potential hadronic $\tau$ background, one can make the stronger requirement that $N_{\rm track} = 2$ or 4.

Finally, we have determined that with the cuts above, it is possible to have high efficiency $Z^\prime$ tagging for a range of $Z'$ masses and couplings. Fig.~\ref{fig:jetsub2} shows the tagging efficiency as a function of the $Z'$ mass, where the shaded band shows the variation among different assumptions on the structure of the $Z'$ coupling to light quarks. In particular, we vary among $Z^\prime$ with isospin-violating vector or axial coupling to light quarks, or vector coupling to only up quarks. The methods described have good efficiency up to $M_{Z'}$ of a few GeV, with the primary difference being the charged particle multiplicity in the $Z'$ jet. In addition, the right panel of Fig.~\ref{fig:jetsub2} shows the $p_T$ dependence for $Z'$ tagging, which is shown to be relatively stable over much of the  $p_T$ range relevant to the LHC. In the rest of the paper, we will adopt a conservative approach and simply assume a constant signal tag efficiency of 50\% and a constant background mistag efficiency of 2\%.

\section{Secluded Dark $Z^\prime$ Model}
\label{sec:secluded}
How the dark matter is produced at colliders is model-dependent, but generically there are two possibilities. The first possibility, which we call the ``secluded dark $Z^\prime$ model''~\cite{Pospelov:2007mp}, is that the SM particles are charge-neutral under the dark $U(1)^\prime$, but have additional interactions with dark matter particles. The second possibility, which we call the ``public dark $Z^\prime$ model'', is to have some SM particles also charged under the dark $U(1)^\prime$ gauge symmetry.  We focus on the secluded $Z'$ model in this section.

\subsection{Elastic Dark Matter}
\label{sec:elastic}
In this first class of models, we consider a $Z^\prime$ under which only the dark matter particle is charged. The interactions between the dark matter and SM particles can be independent of the $Z^\prime$ and described by effective higher dimensional operators. For simplicity, we choose the dark matter particle to be a vector-like Dirac fermion under the $U(1)^\prime$ with a unit charge and an interaction $g_\chi Z^\prime_\mu \overline{\chi} \gamma^\mu \chi$. Concentrating on the up quark, we  consider two effective operators for dark matter production:
\beqa
{\cal O}_V = \frac{\overline{\chi} \gamma^\mu \chi \, \overline{u} \gamma_\mu u }{ \Lambda^2 } \,,   \qquad  \qquad
{\cal O}_A = \frac{\overline{\chi} \gamma^\mu \gamma^5 \chi \, \overline{u} \gamma_\mu \gamma^5 u }{ \Lambda^2 } \,. 
\label{eq:elastic-operator}
\eeqa
The effective dark matter interactions with other quarks can be studied in a similar manner. 

\begin{figure}[t!]
\begin{center}
\includegraphics[width=0.48\textwidth]{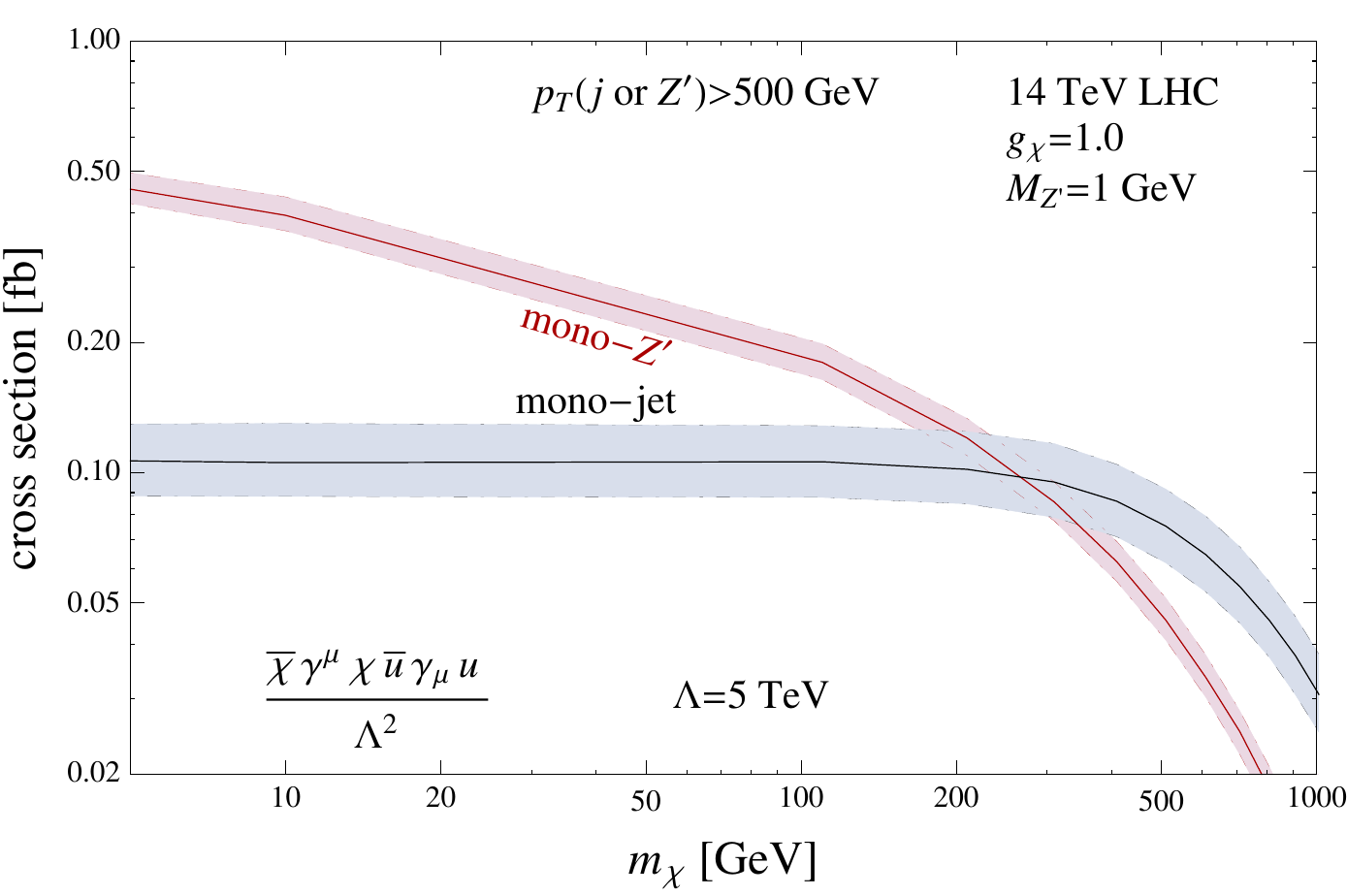}
\includegraphics[width=0.48\textwidth]{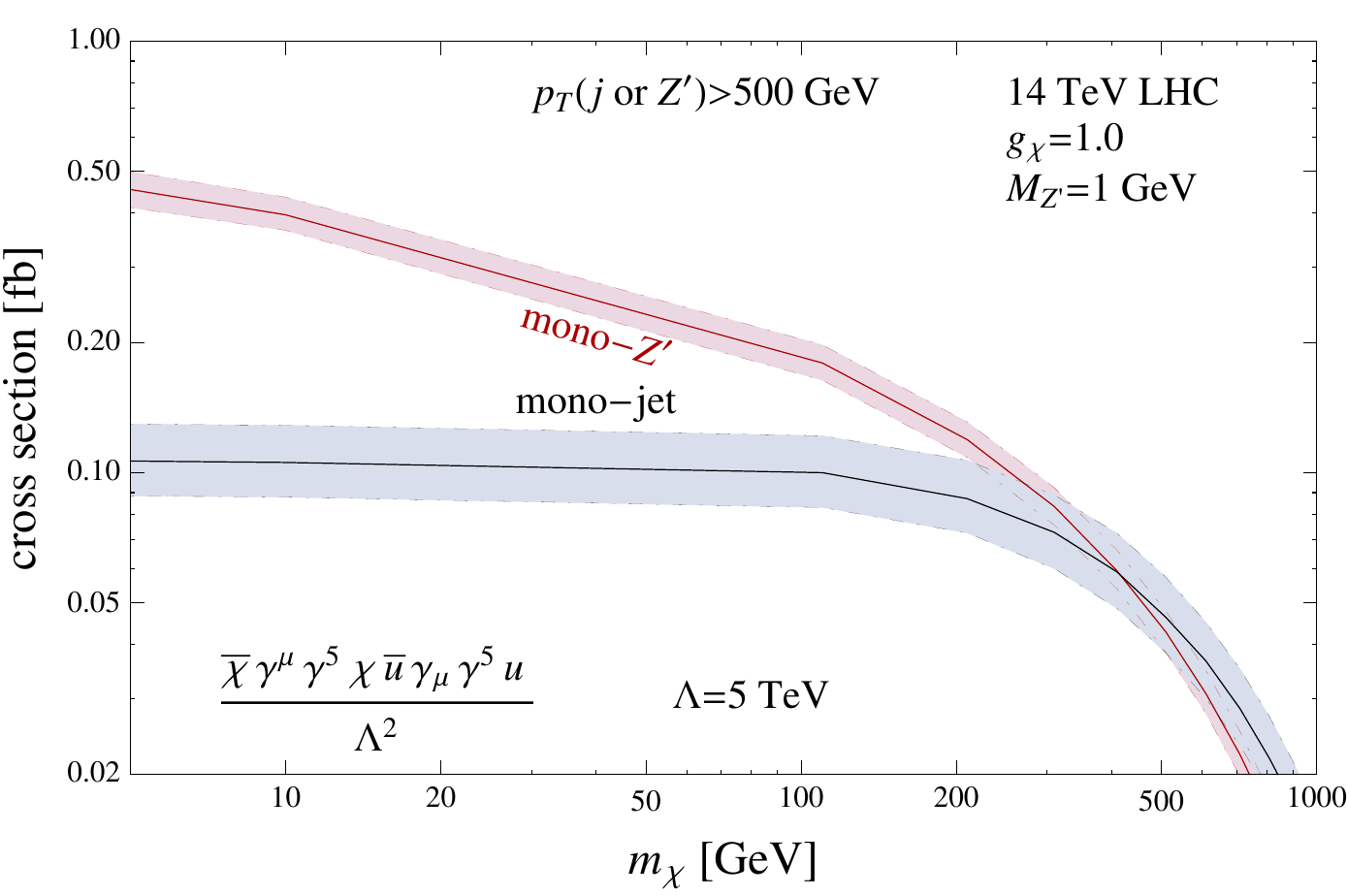}
\caption{Production cross sections of mono-$Z^\prime$ and mono-jet events at the 14 TeV LHC, generated from \texttt{MadGraph}~\cite{Alwall:2014hca}. The band shows the uncertainties from changing the renormalization and factorization scale by a factor of two.}
\label{fig:secluded-zprime-prod}
\end{center}
\end{figure}

For this secluded dark $Z^\prime$ model, the main production of mono-$Z^\prime$ events is from dark matter final state radiation. In Fig.~\ref{fig:secluded-zprime-prod}, we show the production cross sections at the 14 TeV LHC for a light $Z^\prime$ with $g_\chi =1$, $M_{Z^\prime} =1$~GeV and taking a large cutoff $\Lambda=5$~TeV such that we have an approximately valid effective operator description. As a comparison, we also show the ordinary mono-jet production cross sections for the same operator. As one can see, for dark matter mass below around 200 GeV, the mono-$Z^\prime$ production cross section is larger than the mono-jet one for the same dark matter mass. 

\begin{figure}[t!]
\begin{center}
\includegraphics[width=0.48\textwidth]{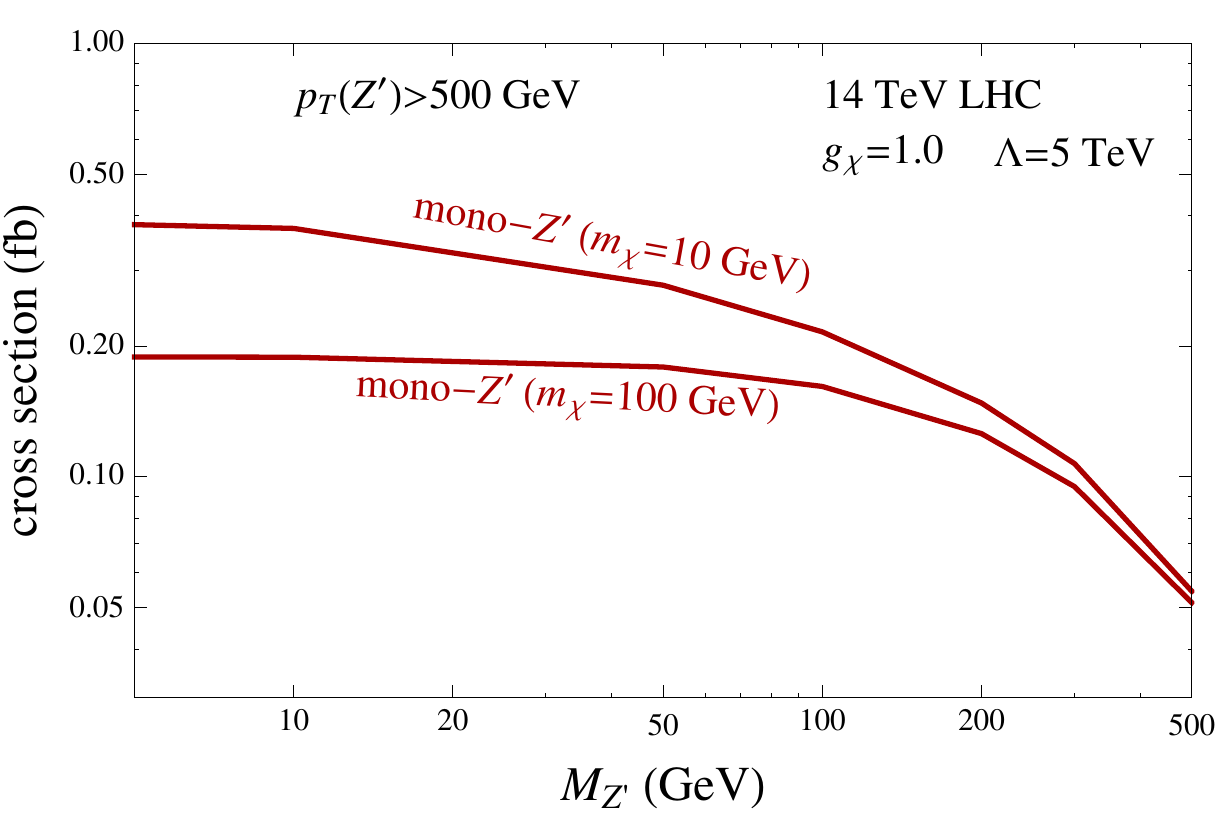}
\caption{Production cross sections of mono-$Z^\prime$ at the 14 TeV LHC; for the masses shown, the result is nearly the same for both the vector and axi-vector interactions. }
\label{fig:secluded-zprime-prod-zprimemass}
\end{center}
\end{figure}

The mono-$Z^\prime$ cross section in Fig.~\ref{fig:secluded-zprime-prod} decreases as the dark matter mass increases even for a light $Z'$ mass. This can be understood by looking at the off-shell dark matter propagator. For final state radiation, $\chi_{\rm off-shell} \rightarrow \chi + Z^\prime$, we can consider the kinematics of the region where $\chi$ and $Z^\prime$ have the same direction in the central direction. So, one has $p(\chi) = \left( \sqrt{(p^{\chi}_T)^2 + m_\chi^2}, p^\chi_T, 0, 0 \right)$ and $p(Z^\prime) = \left( \sqrt{(p^{Z^\prime}_T)^2 + M_{Z^\prime}^2}, p^{Z^\prime}_T, 0, 0 \right)$. The denominator of the off-shell dark matter propagator is
\beqa
[p(\chi)+p(Z^\prime)]^2 - m_\chi^2 \approx \frac{ p^{Z^\prime}_T}{p^{\chi}_T} m_\chi^2 + \frac{p^{\chi}_T}{p^{Z^\prime}_T} M_{Z^\prime}^2 + M_{Z^\prime}^2     \,,
\eeqa
in the limit of $p^{Z^\prime}_T, p^{\chi}_T  \gg m_\chi, M_{Z^\prime}$. Since large $p_T^{Z^\prime} = E_T^{\rm miss}$ is required for the mono-$Z'$ event, both the dark matter mass  and the $Z'$ mass must be relatively light in order to boost the rate. Therefore, as we see above, a smaller dark matter mass leads to an increase in the production cross section. Similarly, the cross section is larger for a lighter $Z'$:  in Fig.~\ref{fig:secluded-zprime-prod-zprimemass}, we show the mono-$Z^\prime$ production cross sections as a function of the $Z^\prime$ mass for a fixed dark matter mass. 

If the dark $Z^\prime$ mass is above twice of the dark matter mass, it will mainly decay into two dark matter particles, which are invisible at colliders. One can rely on the standard mono-jet events to discover this scenario. On the other hand, for $M_{Z^\prime} \leq 2 m_\chi$ the $Z^\prime$ can only decay back to the SM particles via the higher-dimensional operators in Eq.~(\ref{eq:elastic-operator}). For a modestly large cutoff, the lifetime of $Z^\prime$ could be sufficiently long to have a displaced-vertex collider signature. We next calculate the dark $Z^\prime$ lifetime for both higher-dimensional operators.

For the vector-like coupling operator ${\cal O}_V$, the effective coupling between $Z^\prime$ and up quark can be described by the following current-current operator~\cite{Fox:2011qd}
\beqa
 \frac{\tilde{c}}{\Lambda^2}\, (\phi^{\prime\,\dagger} D_\mu \phi^\prime - \phi^{\prime} D_\mu \phi^{\prime\,\dagger})\,(\overline{u}\gamma^\mu u) \,,
 \label{eq:decay-operator}
\eeqa
in the unbroken $U(1)^\prime$ theory. Here, the parameter $\tilde{c}$ is introduced to indicate the unknown UV parameter and $\phi^\prime$ is the scalar field that develops a vacuum expectation value (VEV), $\langle\phi^\prime\rangle=v^\prime/\sqrt{2}$, to spontaneously break the $U(1)^\prime$ symmetry. One example of having $\tilde{c} = O(1)$ is to introduce another massive $Z^{\prime \prime}$ for generating the effective operator ${\cal O}_V$. If the scalar $\phi^\prime$ is also charged under $Z^{\prime \prime}$, the operator in Eq.~(\ref{eq:decay-operator}) can be generated at tree-level by integrating out $Z^{\prime \prime}$. Another example is to have the kinetic mixing parameter between $U(1)^\prime$ and $U(1)^{\prime\prime}$, which could be loop-factor-suppressed if it just comes from the dark matter loop. 

If the VEV of the $U(1)^\prime$-charged scalar field $\phi^\prime$ is zero or the $U(1)^\prime$ is unbroken, the effective charge coupling of $Z^\prime_\mu \, \overline{u} \gamma^\mu u$ is zero. This is a manifestation of well-known fact in the literature that particles charged under a massive gauge boson (the heavy mediator $Z^{\prime\prime}$ to generate the effective operator) will not have a millicharge under the unbroken massless gauge boson~\cite{BH1986}. On the other hand, for a nonzero VEV $\langle \phi^\prime \rangle = v^\prime/\sqrt{2}$ and a massive $Z^\prime$, the effective coupling becomes
\beqa
c'\,\frac{M_{Z^\prime}^2}{\Lambda^2}\,Z^\prime_\mu \, \overline{u} \gamma^\mu u \,.
\eeqa
Here, the parameter $c'$ is related to $\tilde{c}$ by the $U(1)^\prime$ gauge coupling. 

For the interesting parameter space with $M_{Z^\prime} =O(1~\mbox{GeV})$, $Z^\prime$ will decay into just a few hadrons and is therefore quite different from an ordinary QCD jet at high $p_T$. We use the chiral Lagrangian to convert the operators in terms of quark fields to the operators in terms of pions: $\overline{u} \gamma_\mu u \rightarrow \pi^+ \partial_\mu \pi^- - \pi^- \partial_\mu \pi^+ + K^+ \partial_\mu K^- - K^- \partial_\mu K^+$. The decay width for $Z^\prime \rightarrow \pi^- \pi^+$ is calculated to be 
\beqa
\Gamma(Z^\prime \rightarrow \pi^- \pi^+) = \frac{M_{Z^\prime}}{48\,\pi} \, \left(\frac{c' \,M^2_{Z^\prime} }{\Lambda^2} \right)^2\,\left(1 - \frac{4\,m_\pi^2}{M_{Z^\prime}^2}   \right)^{3/2} \,.
\eeqa
A similar result can be obtained for $Z^\prime \rightarrow K^- K^+$ by replacing $m_\pi$ with $m_K$, leading to a more suppressed phase space factor. For $c'=1$, $M_{Z^\prime} = 1$~GeV and $\Lambda=1$~TeV, the travel distance of $Z^\prime$ before it decays is
\beqa
c\tau_0 \approx 3~\mbox{cm} \,,
\eeqa
which can be a displaced $Z^\prime$ at the LHC. In our following sensitivity study, we will treat the $Z^\prime$ decay length as a free parameter and mainly concentrate on the prompt decay case.  

For the other axi-vector operator, a similar UV completion model can lead to the following operator for $Z^\prime$ decay
\beqa
d'\,\frac{M_{Z^\prime}^2}{\Lambda^2}\,Z^\prime_\mu \, \overline{u} \gamma^\mu\gamma_5 u \,,
\eeqa
with $d'$ as a dimensionless and model-dependent number. Using the chiral Lagrangian and treating the $\rho$ meson as the gauge boson of a hidden local gauge symmetry~\cite{Fujiwara:1984mp}, we have the operator translation: $\overline{u} \gamma_\mu \gamma_5 u \rightarrow 2 i g_{\rho \pi \pi} f_\pi (\rho^+_\mu \pi^- - \rho^-_\mu \pi^+)$ with $f_\pi \approx 92$~MeV and the $\rho \pi \pi$ coupling $g_{\rho \pi \pi}^2/4\pi \approx 3.0$.  The decay width of $Z^\prime \rightarrow \rho \pi$ is 
\beqa
\Gamma(Z^\prime \rightarrow \rho \pi) =2\,\Gamma(Z^\prime \rightarrow \rho^+ \pi^-) = \frac{d'^2\, g_{\rho \pi \pi}^2 \, f_\pi^2\, M_{Z^\prime}^2\,p}{3\pi \Lambda^4}\, \left( 3 + \frac{p^2}{m_\rho^2}\right)\,,
\eeqa
with $p^2 = [M_{Z^\prime}^2 - (m_\rho + m_\pi)^2] [M_{Z^\prime}^2 - (m_\rho - m_\pi)^2]/4M^2_{Z^\prime}$. For $d'=1$, $M_{Z^\prime} = 1$~GeV and $\Lambda=1$~TeV, the travel distance of $Z^\prime$ before it decays is
\beqa
c\tau_0 \approx 1.2~\mbox{cm} \,,
\eeqa
comparable to the vector-like coupling case. The charged $\rho$-meson will subsequently decay into $\pi^\pm \pi^0$. For both vector and axi-vector cases, the $Z'$ boson decays to two charged hadrons. Therefore, we perform a collider study for this interesting class of {\it mono-$Z^\prime$ jet} signatures, which have a light $Z^\prime$ mainly decaying into  two or three hadrons (with two of the hadrons charged).

There could exist other ways to provide $Z^\prime$ decays which may give different collider signatures. A common one is through the kinetic mixing with the hyper-charge gauge group, $\frac{1}{2}\,Z^{\prime}_{\mu\nu} B^{\mu\nu}$~\cite{BH1986}.  The $Z^\prime$ then can have a substantial decay width into leptons and is easier to be searched for at the LHC (see Ref.~\cite{CMS:2014hka} for the CMS searches for displaced dileptons at 8 TeV with 20 fb$^{-1}$). However, even if the $Z^\prime$ has a sizable branching ratio to leptons, as long as there is a decay to light quarks, then $Z^\prime$-jets are an additional unique signature and can be searched for in complementary channels.

\subsubsection{Sensitivity at the 14 TeV LHC}
\label{sec:sensitivity}

For the traditional searches for dark matter in the mono-jet channel, the dominant background is production of weak gauge bosons plus multiple jets.  To estimate the constraints on our dark matter models, we use \texttt{FeynRules}~\cite{Alloul:2013bka} to create a model file for \texttt{MadGraph}~\cite{Alwall:2014hca} and then generate events at parton level. We then use \texttt{Pythia}~\cite{Sjostrand:2007gs} to shower and hadronize the parton-level events. Finally, we use \texttt{PGS}~\cite{PGS} to cluster hadrons into jets as well as to perform the detector simulation. The existing searches at the 8 TeV LHC have shown a large systematic uncertainty due to jet energy scale, PDF's and so on~\cite{Khachatryan:2014rra}. Using the jet-substructure cuts to further suppress background events, we anticipate a significant improvement on top of the ordinary mono-jet analysis. 

\begin{figure}[t!]
\begin{center}
\includegraphics[width=0.46\textwidth]{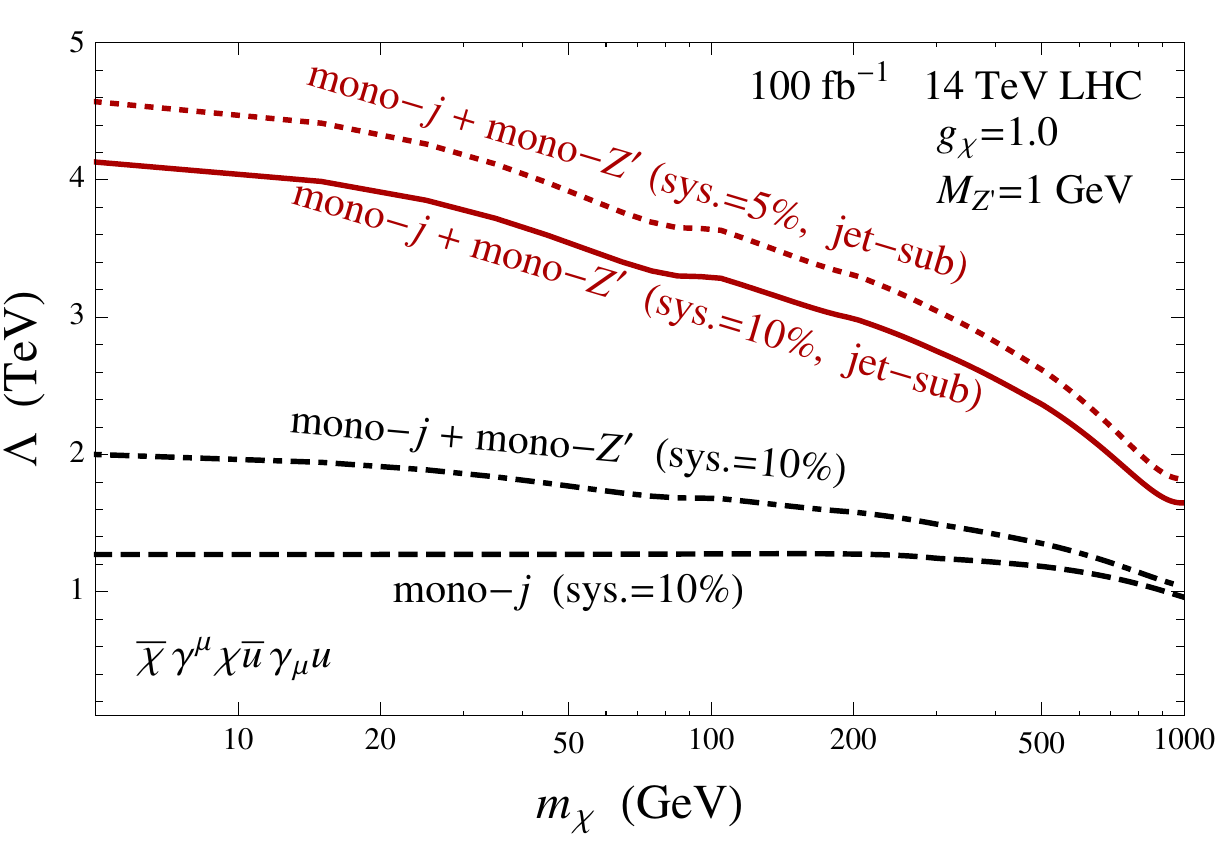} \hspace{3mm}
\includegraphics[width=0.46\textwidth]{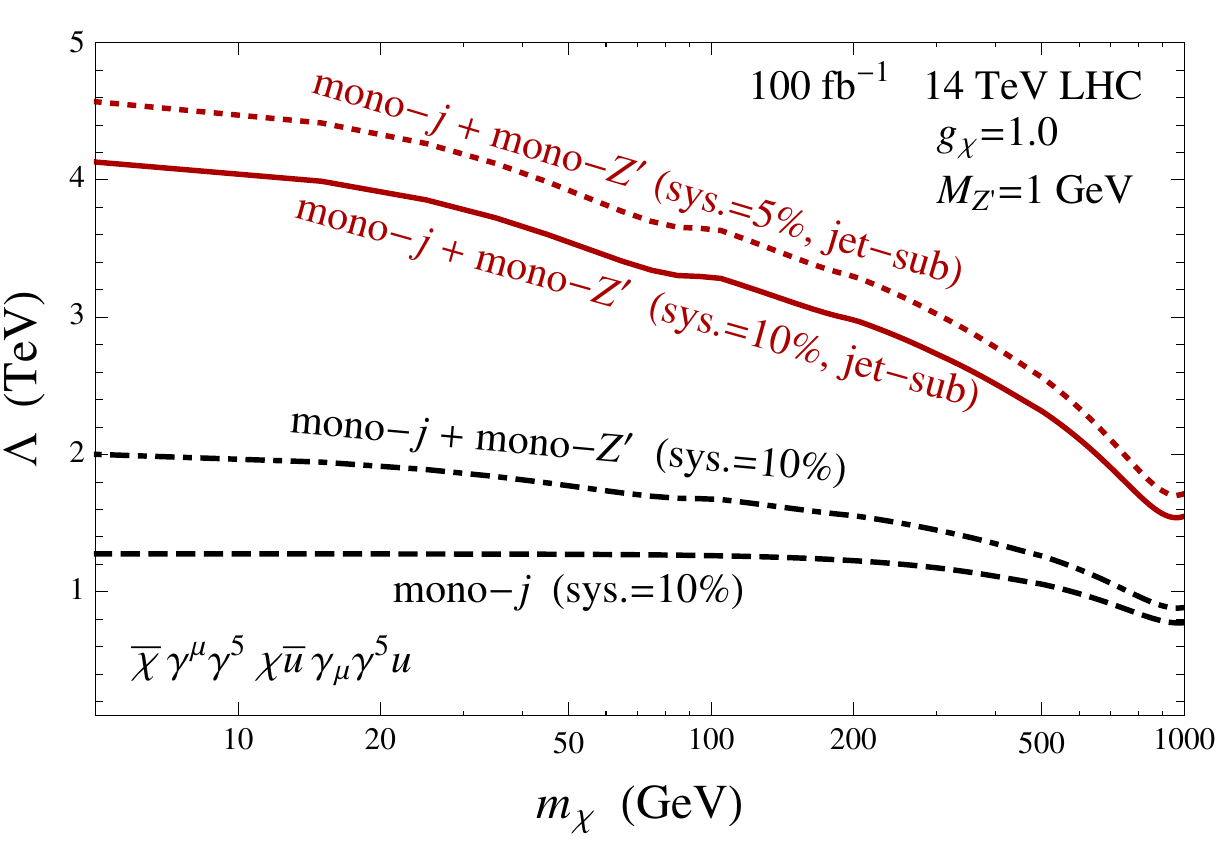}
\caption{Left panel: the 90\% C.L. constraints on the cutoff for the vector-vector interaction. A cut on the missing energy is taken to be $E_T^{\rm miss} > 500$~GeV. The black and lower lines are using the traditional mono-jet analysis. The red and upper lines are based on a jet-substructure analysis for the mono-$Z^\prime$ jet. The systematic error is assumed to be 10\% (5\%) for the solid (dashed) line. The mono-$Z^\prime$ tag efficiency is chosen to be 50\% and the background mistag efficiency is 2\%. Right panel: the same as the left one but for the axi-vector interaction operator.}
\label{fig:cutoff-constraints}
\end{center}
\end{figure}

Guided by the CMS search for the mono-jet signature at the 8 TeV LHC~\cite{Khachatryan:2014rra}, we fix the most important cut to be $E_T^{\rm miss} > 500$~GeV. We require the leading jet to have $p_T$ above 200~GeV and veto events with a second jet with $p_T>60$~GeV or with a lepton with $p_T>20$~GeV. The total background cross section, $W/Z$ + jets, is simulated to be 142~fb at leading order and after applying detector-level cuts.  For the mono-$Z^\prime$ signature, there is also a potential background from $W(\tau \nu)$+jets with hadronic-$\tau$ decay. We neglect this background, since it can be suppressed by a veto on $\tau_h$. The total cross section for  $W(\tau_h \nu)$+jets is 1.1 fb after the $E_T^{\rm miss} >$ 500~GeV cut and 0.8 fb after imposing our $Z^\prime$-jet tagging. An additional veto on events with one or three tracks in the inner $\Delta R=0.2$ "isolation cone'' can reduce this background to 0.05 fb, which is negligible.

In Fig.~\ref{fig:cutoff-constraints}, we show the 90\% C.L. constraints on the cutoff of the effective operators for two different assumptions of systematic errors. Compared to the current limits, which give $\Lambda \gtrsim 1030$~GeV for light dark matter at the 8 TeV LHC, the 14 TeV LHC mono-jet analysis can improve the limits by around 30\%. By including the contribution of FSR from mono-$Z'$, the limits can be improved by another 50$\%$. On the other hand, if one performs a dedicated analysis for the mono-$Z^\prime$ jet signature, the constraints on the cutoff can be improved by another factor of two. This is because the current mono-jet searches are limited by the systematic errors and the unique characteristics of the mono-$Z^\prime$ jet can dramatically reduce the background events.

For a long-lived $Z^\prime$, the signature is so peculiar such that the SM background is expected to be negligibly small. The existing search on displaced dijets has focused on heavier particle masses above 50 GeV~\cite{CMS:2014wda,Han:2007ae}. The light $Z^\prime$ should behave more like a $\tau$-lepton, with smaller vertex track multiplicity and smaller jet mass. We do not perform a detailed analysis for the long-lived $Z^\prime$ case in this paper. For light dark matter and still requiring $p_T(Z^\prime) > 500$~GeV, one can obtain a constraint on the cutoff as large as $\Lambda > 8$ TeV by assuming negligible backgrounds and allowing up to five signal events. A dedicated analysis with a less stringent $p_T$ cut on $Z^\prime$ is very likely to have even better sensitivity.

\begin{figure}[t!]
\begin{center}
\includegraphics[width=0.46\textwidth]{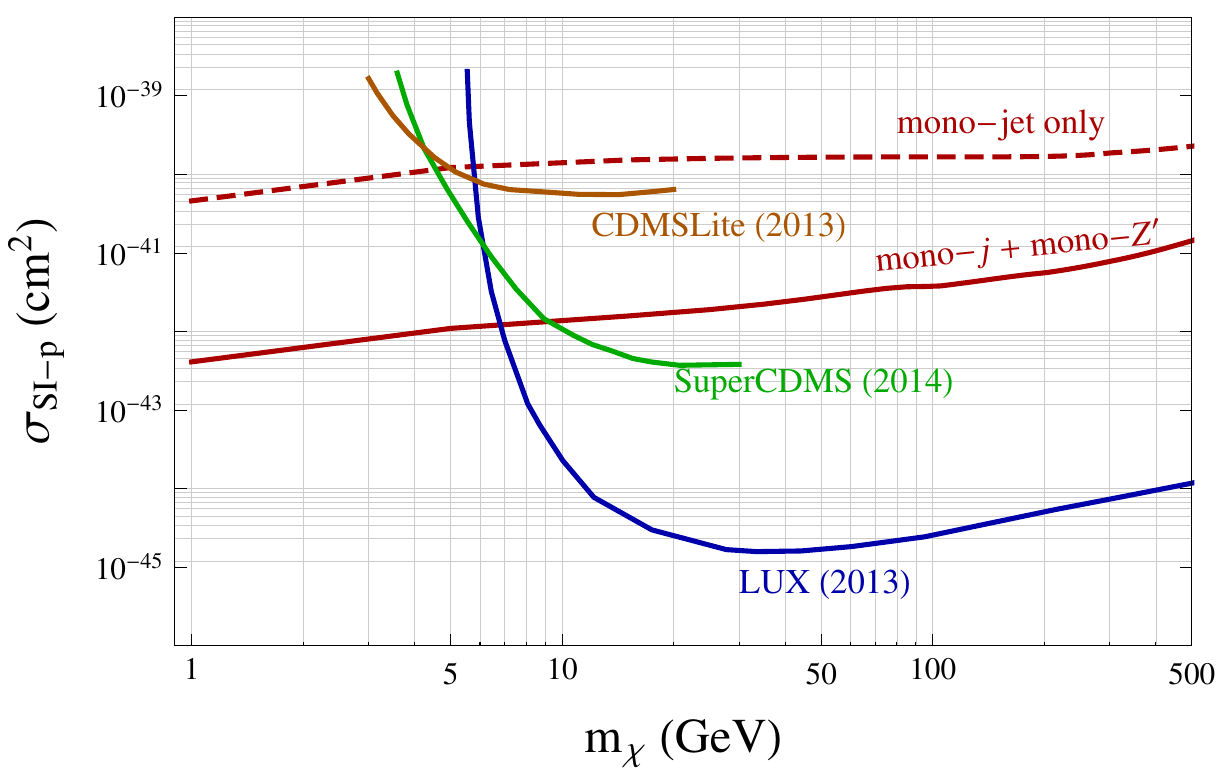} \hspace{3mm}
\includegraphics[width=0.46\textwidth]{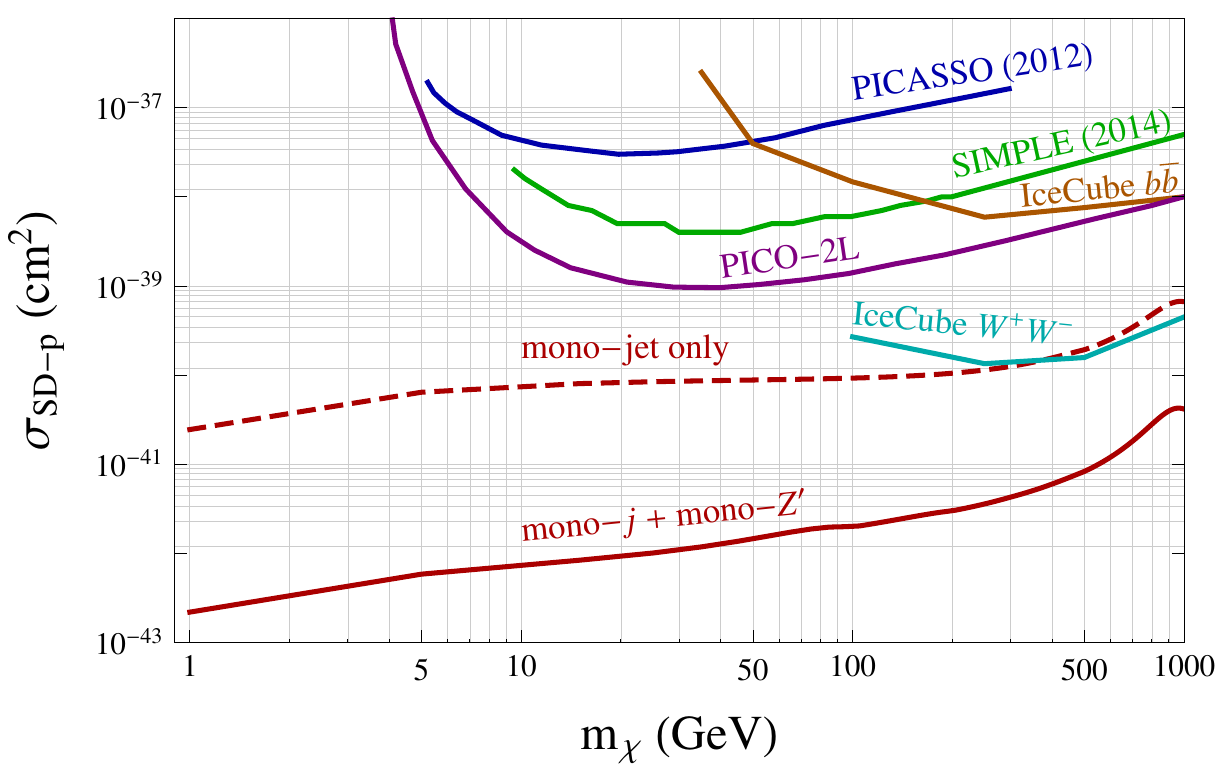}
\caption{Left panel: projected constraints on dark matter-proton spin-independent scattering cross sections from the standard mono-jet analysis at the 14 TeV LHC with 100 fb$^{-1}$ and the mono-$Z^\prime$ jet-substructure based analysis. The model parameters are $M_{Z^\prime}=1$~GeV and $g_\chi=1.0$, and we take the limits on $\Lambda$ assuming $10\%$ systematic error.  Also shown are the current constraints from direct detection experiments: LUX~\cite{Akerib:2013tjd}, SuperCDMS~\cite{SuperCDMS} and CDMSLite~\cite{CDMSLite}. Right panel: similar to the left panel but for dark matter-proton spin-dependent scattering cross sections. The current experimental bounds are from: PICASSO~\cite{Archambault:2012pm}, SIMPLE~\cite{Felizardo:2014awa}, PICO-2L~\cite{Amole:2015lsj} and IceCube~\cite{Aartsen:2012kia}. }
\label{fig:directdetection}
\end{center}
\end{figure}

To compare to other limits from direct detection experiments, we convert the limits on the cutoff from colliders into dark matter-nucleon scattering cross sections~\cite{Bai:2010hh}. Since we have only considered the example of an operator coupling to the up-quark, the $\chi$-proton and $\chi$-neutron spin-independent scattering cross sections are different. We therefore scale the limits from spin-independent direct detection experiments by a factor of $4 A^2/(A+Z)^2$ and show them in the left panel of Fig.~\ref{fig:directdetection}. Although the jet-substructure analysis from the mono-$Z^\prime$ can dramatically increase the sensitivity, the direct detection experiments still provide the best limit for  dark matter mass above 6 GeV. For lighter dark matter mass, the collider will eventually provide the best limit. In the right panel of Fig.~\ref{fig:directdetection}, we show the limits on the spin-dependent scattering cross sections. As one can see, the collider will provide the best limits for a wide range of masses until around 1 TeV even without considering the mono-$Z^\prime$ signal. Under the assumptions above, the mono-$Z^\prime$ signature will significantly enhance the discovery potential and easily compete with a next-generation spin-dependent dark matter experiment such as PICO. 

Finally, we note that in mapping the sensitivity for the cutoff scale $\Lambda$ onto the direct detection plane, the contribution from $Z^\prime$-mediated nucleon scattering has been neglected. For a GeV-scale $Z^\prime$  with couplings to quarks $\gtrsim 10^{-5}$, such that the $Z^\prime$ decay is prompt on collider scales, then the scattering rate through the $Z^\prime$ may be much larger than the quoted collider bound. For example, if the $Z^\prime$ has vector interactions with the dark matter and quarks, then $\sigma_{\textrm{SI-p}} \sim 10^{-40}$ cm$^2$. However, the direct detection cross section depends on the specifics of the $Z^\prime$ couplings and could also be velocity-suppressed, so we do not include this.

\subsection{Inelastic Dark Matter}
\label{sec:idm}
Next we consider a dark matter sector with an inelastic splitting between the ground state $\chi$ and excited state $\chi_*$. The kinematics of the mono-$Z'$ signal is now different if the decay $\chi_* \to \chi Z'$ is permitted. We introduce Dirac dark matter fields with an off-diagonal coupling to $Z^\prime$ as
\beqa
g_\chi\, \left(\overline{\chi}_* \gamma^\mu \chi  + \overline{\chi} \gamma^\mu \chi_* \right) \, Z^\prime_\mu \,.
\label{eq:idm-coupling}
\eeqa
A simple way to realize this interaction without any corresponding diagonal interactions is to have two Dirac fermions, $\chi_1$ and $\chi_2$, which have opposite charges under the $U(1)^\prime$ symmetry but identical masses. The choice of equal masses and opposite charges is protected by a matter parity under which: $\chi_1 \rightarrow - i \gamma^2 \chi_2^*$ and $\chi_2 \rightarrow - i \gamma^2 \chi_1^*$. To generate a mass splitting between those two states, one can introduce the matter parity breaking operator $\lambda\,\phi^\prime \bar{\chi}_1 \chi_2 + \lambda^*\,\phi^{\prime\,\dagger} \bar{\chi}_2 \chi_1$. Here, the scalar field $\phi^\prime$ has a non-zero VEV to break the $U(1)^\prime$ gauge symmetry. Rotating to the mass eigenstate, $\chi = (\chi_1 - \chi_2)/\sqrt{2}$ and $\chi_* = (\chi_1 + \chi_2)/\sqrt{2}$, we have only the off-diagonal coupling in Eq.~(\ref{eq:idm-coupling}). For this specific realization, we anticipate the mass difference $\Delta \equiv m_{\chi_*} - m_\chi$ to be at the same order of magnitude as $M_{Z^\prime}$ and can be dramatically smaller than the dark matter mass. 

The main decay channel of $\chi_*$ is $\chi_* \rightarrow \chi + Z^\prime$ and has the decay width
\beqa
\Gamma(\chi_* \rightarrow \chi + Z^\prime) &=& \frac{g_\chi^2}{16\pi m_{\chi_*}^3 M_{Z^\prime}^2} 
\left[ ( m_{\chi} + m_{\chi_*})^2 + 2 M_{Z^\prime}^2  \right] \,\left[ (m_{\chi_*} - m_\chi)^2 - M_{Z^\prime}^2 \right] \nonumber \\
&&  \times \, \sqrt{ \left[ (m_{\chi_*} + m_\chi)^2 - M_{Z^\prime}^2  \right] \left[ (m_{\chi_*} - m_\chi)^2 - M_{Z^\prime}^2  \right]   } \nonumber \\
&\approx& \frac{g_\chi^2}{2\pi\,M_{Z^\prime}^2}\,(\Delta^2 - M_{Z^\prime}^2)^{3/2}\,.
\eeqa
where in the second line we have taken the limit of $\Delta \sim M_{Z^\prime} \ll m_\chi$. For $g_\chi=1$, $\Delta =2$~GeV and $M_{Z^\prime}=1$~GeV, we have $\Gamma(\chi_* \rightarrow \chi + Z^\prime) = 0.83$~GeV for a very heavy dark matter mass. Similar to the elastic dark matter model, there can be an interaction $Z^\prime_\mu \bar{u} \gamma^\mu u$ which allows the light $Z^\prime$ to decay into two or more charged hadrons.

\begin{figure}[t!]
\begin{center}
\includegraphics[width=0.48\textwidth]{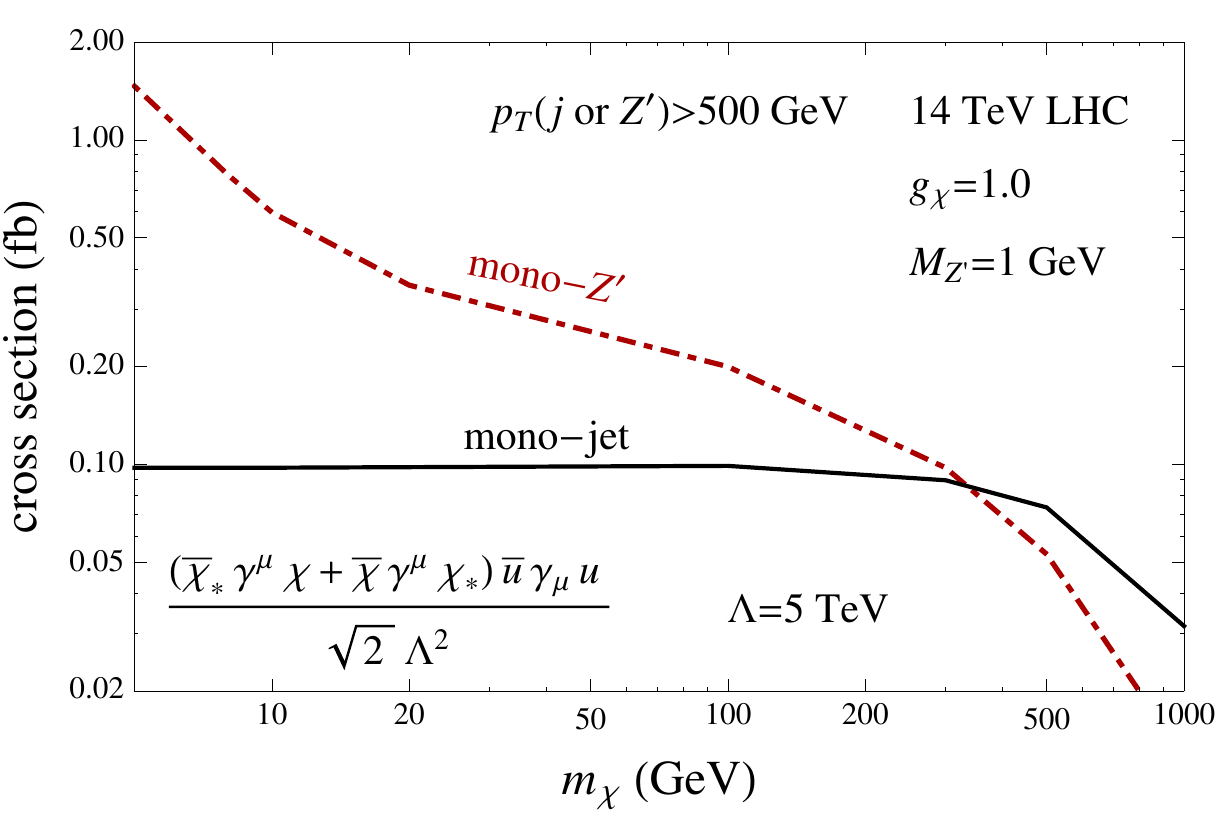}
\includegraphics[width=0.465\textwidth]{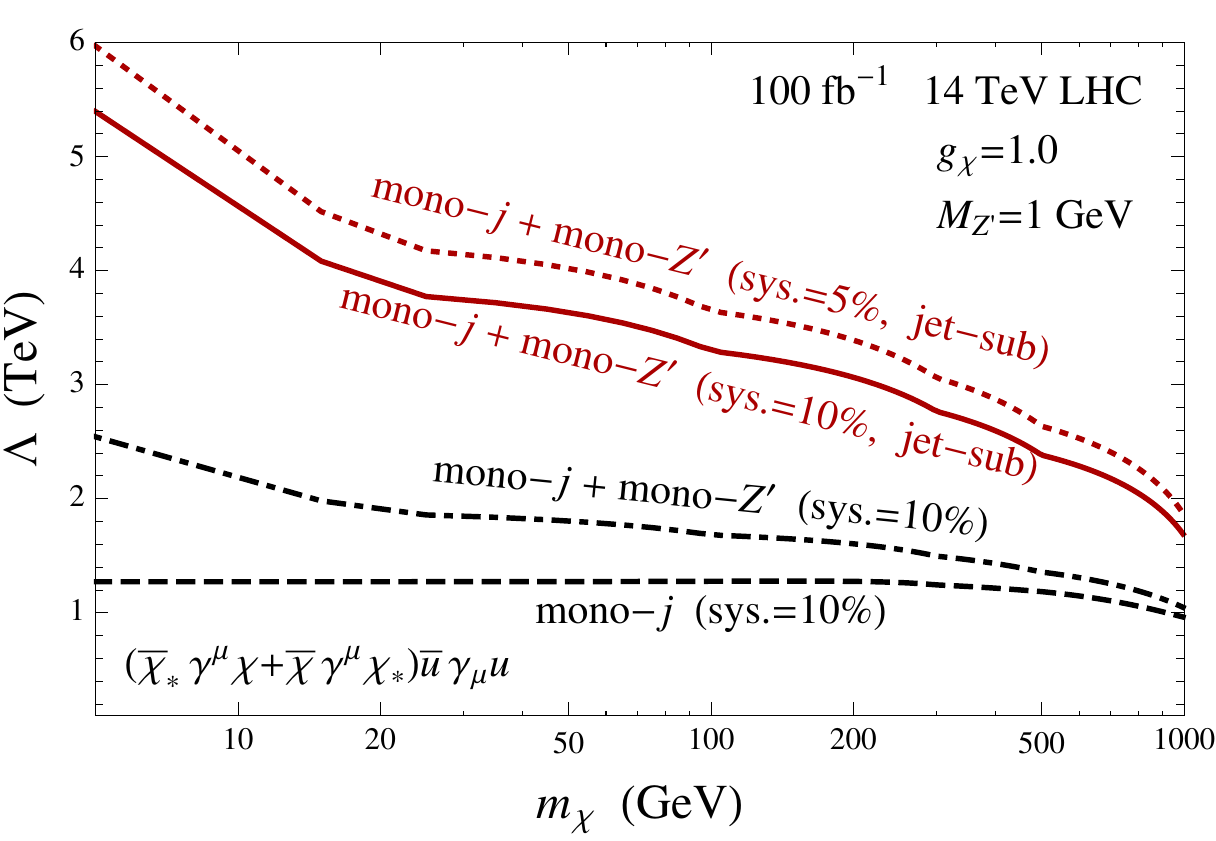}
\caption{Left panel: production cross sections of mono-$Z^\prime$ in the iDM model with a contact interaction and mass splitting $\Delta = 2$ GeV. Right panel:  the 90\% C.L. constraints on the cutoff for the vector-vector interaction. For the jet-substructure analysis, we choose 50\% for the signal tag efficiency and 2\% for the background mistag efficiency.}
\label{fig:secluded-zprime-prod-idm}
\end{center}
\end{figure}

We introduce effective higher-dimensional operators to couple dark matter to the SM quarks. As an example, we consider the following operator
\beqa
\frac{(\overline{\chi}_* \gamma^\mu \chi + \overline{\chi} \gamma^\mu \chi_*)\, \overline{u} \gamma_\mu u }{\sqrt{2}\, \Lambda^2 } \,, 
\label{eq:inelastic-opp}
\eeqa
where we introduce the factor of $1/\sqrt{2}$ to have the same mono-jet production for the same cutoff defined in Eq.~(\ref{eq:elastic-operator}). At the LHC, the signal process contains both the two-body production with a subsequent decay, $pp \rightarrow \chi \overline{\chi}_* \rightarrow \chi \overline{\chi} Z^\prime$, or a three-body production $pp \rightarrow \chi \overline{\chi} Z^\prime$. For a stringent cut like $p_T(Z^\prime) > 500$~GeV, the two-body productions become important only for a light dark matter mass because of the need of a large boost for $Z^\prime$. If the excited state is light enough, then it can be produced on-shell with a large boost $O(100)$, which in turn allows the $Z'$ produced in the decay to have a large enough momentum $p_T(Z^\prime) = O(100 M_{Z^\prime})$ and pass the $p_T$ cut. For heavy dark matter mass, the three-body process dominates and the small mass splitting of the two states can be ignored, such that the production cross sections follow the same behavior as in the elastic case. For $\Lambda=5$~TeV, we show the production cross section at the 14 TeV LHC in the left panel of Fig.~\ref{fig:secluded-zprime-prod-idm}. As one can see, for dark matter mass below roughly 10 GeV there is an enhanced production cross section due to the two-body process. In the right panel of Fig.~\ref{fig:secluded-zprime-prod-idm}, we show the potential constraints on the cutoff from the standard mono-jet analysis and the jet-substructure based mono-$Z^\prime$ analysis. Comparing to the limits for the elastic dark matter model in Fig.~\ref{fig:cutoff-constraints}, one can see that for the inelastic dark matter model the constraints for light dark matter below 10 GeV are more stringent.

\section{Public Dark $Z^\prime$ Model}
\label{sec:public}
In this subsection, we discuss a class of dark $Z^\prime$ models with both dark matter and some SM fermions charged under $U(1)^\prime$. For this class of models, the relevant couplings include $g_q$ and $g_\chi$, of the $Z^\prime$ couplings to quarks and dark matter. For a heavy $Z^\prime$ where the $Z^\prime$ can decay invisibly, the current mono-jet search constrains the effective cut-off $\Lambda \equiv M_{Z^\prime}/ \sqrt{g_q g_\chi}$ to be above around 1 TeV for dark matter mass below 200 GeV~\cite{Khachatryan:2014rra,Aad:2015zva}. For couplings of order of unity, the $Z^\prime$ mass is therefore constrained to be very heavy, especially when the $Z^\prime$ can be  produced on-shell at the LHC. For this region of parameter space, it is challenging to have sufficient mono-$Z^\prime$ events either from ISR or from FSR. On the other hand, the mono-jet constraints no longer apply for a light $Z^\prime$, or if the $Z^\prime$ can no longer decay to the dark matter. This again motivates the study of a light $Z^\prime$ and mono-$Z^\prime$ jets.

The $U(1)^\prime$ gauge coupling is significantly constrained by direct detection experiments: if the dark matter mass is above $O(5~\mbox{GeV})$, LUX~\cite{Akerib:2013tjd} limits dominate, and below $O(5~\mbox{GeV})$, the best limits are from Xenon10 and CRESST-II~\cite{Angle:2011th,Angloher:2014myn}. For vector-like couplings of $Z^\prime$ to quarks and dark matter particles, we have the constraint of $g_q g_\chi \lesssim 3 \times 10^{-5}$ for $m_\chi = M_{Z^\prime}=2$~GeV from CRESST-II~\cite{Angloher:2014myn}. This means that the allowed gauge coupling is tiny and that the probability to emit a $Z^\prime$ from FSR is very small.

The situation is changed if the dark matter mass is below $\sim 1$~GeV. At the CRESST-II experiment~\cite{Angloher:2014myn}, the lower limit of accepted recoil energies is 0.6~keV and is above the typical recoil energies from a 1 GeV dark matter particle. The constraints from direct detection are much weaker and a sizable value of $g_q g_\chi$ may be still allowed. For a chiral dark matter particle charged under the new $U(1)^\prime$ symmetry, the dark matter mass should be bounded roughly by the $Z^\prime$ mass for perturbative Yukawa couplings. Therefore, in this section we concentrate on the parameter space with a light dark matter particle and a light $Z^\prime$.

For SM particles charged under the dark $U(1)^\prime$, the immediate constraint comes from gauge anomaly cancellation. In particular, for production at the LHC we require some flavors of light quarks to be charged under the $U(1)^\prime$, while we avoid charges for the first and second generation leptons since these are extremely constrained. One way for this to be anomaly free is to introduce new fermions chiral under $SU(2)_W\times U(1)_Y$, but then either there are very stringent constraints from the $Z$ boson decay or the fermions have a mass that is too large compared to the light $Z^\prime$ mass (see discussion and constraints in Refs.~\cite{Carone:1994aa,Carone:1995pu,Graesser:2011vj,Dobrescu:2014fca,Tulin:2014tya} for a baryonic GeV-scale $Z^\prime$). Without any new fermion chiral under the electroweak symmetry, one can instead have generation-dependent charges. One possibility that includes a chiral dark matter particle under the $U(1)^\prime$ is the charge assignment
\beqa
z_{u_R} = 1 \,, \quad  z_{d_R} = -1 \,,  \quad z_{\tau_R} = -1\,,  \quad z_{\chi_R} = 1\,,  \quad z_{\chi_L} = 0 \,,
\eeqa
with other fermions neutral under $U(1)^\prime$. This can be extended to allow a vector-like dark matter particle under $U(1)^\prime$,
\beqa
z_{u_R} = 1 \,, \quad  z_{d_R} = -1 \,,  \quad z_{\tau_R} = -1\,,  \quad z_{\psi_R} = 1 \,, \quad z_{\psi_L} = 0 \,,\quad z_{\chi_R} = r\,,  \quad z_{\chi_L} = r \,,
\eeqa
where we have introduced a new fermion $\psi$ in the dark matter sector. Depending on the relative charges of the dark matter and SM particles, we can have different interaction strengths of the $Z^\prime$ with dark matter and quarks.

\begin{figure}[t!]
\begin{center}
\includegraphics[width=0.48\textwidth]{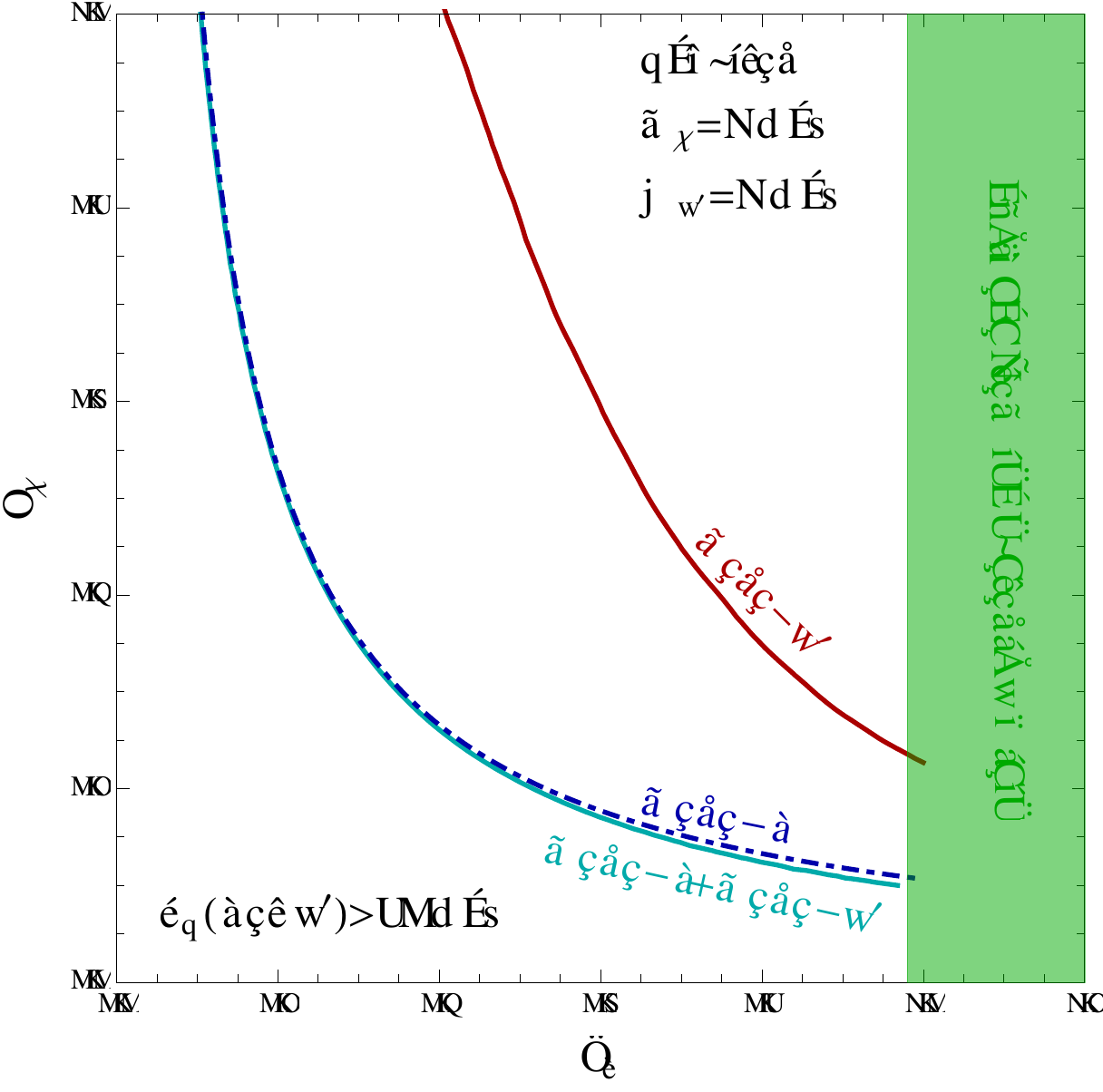}
\includegraphics[width=0.48\textwidth]{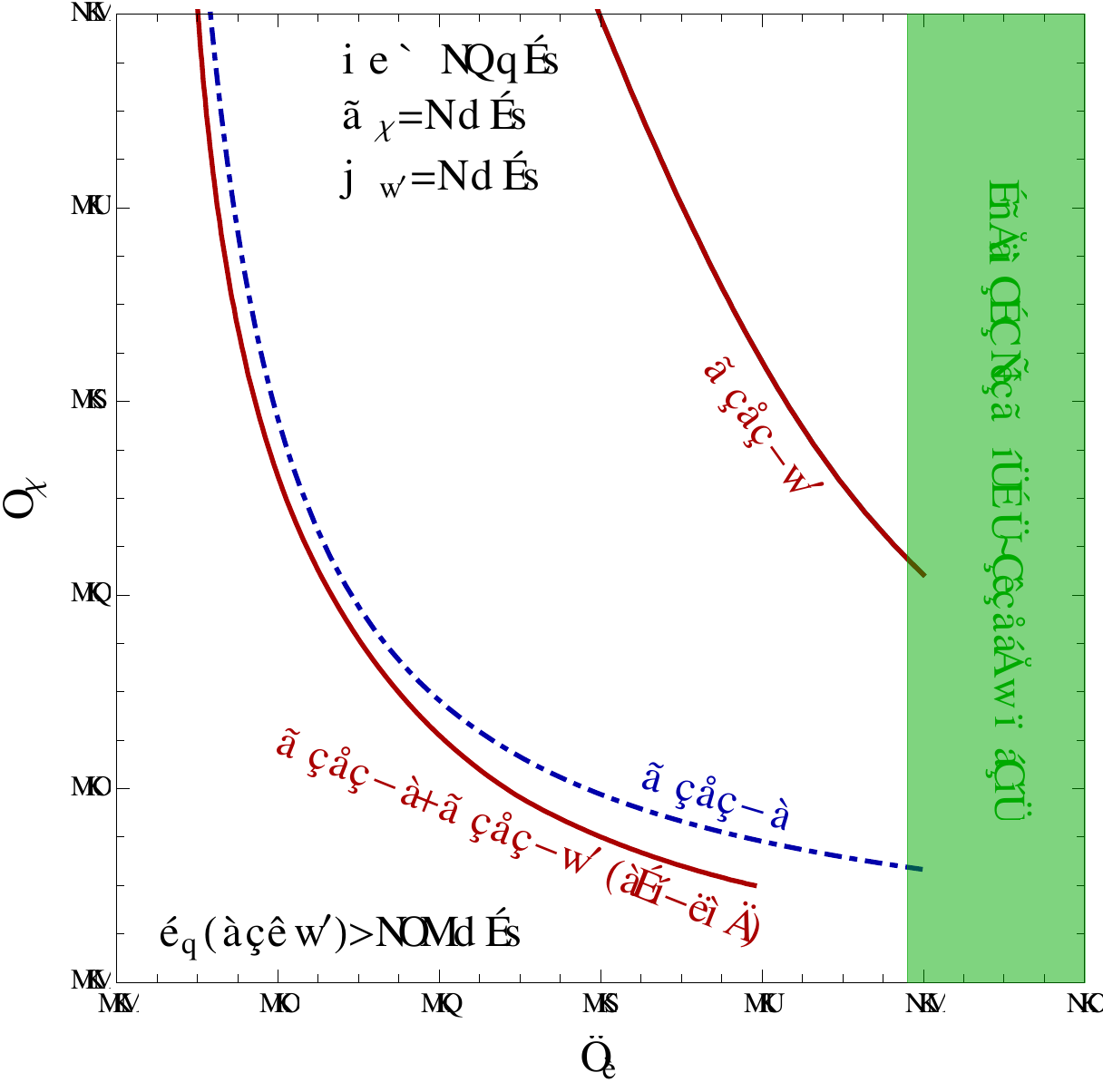}
\caption{Left panel: the 90\% C.L. constraints on the light $Z^\prime$ couplings from the Tevatron~\cite{tevatron-monojet-link} with 1.96~TeV and 1.0 fb$^{-1}$. The shaded region is excluded by the $Z$ boson hadronic width at 90\% C.L. Right panel: the projected sensitivity at the 14 TeV LHC with 100 fb$^{-1}$. The systematic error is assumed to be 5\%.
\label{fig:constrain-tevatron}  
}
\end{center}
\end{figure}

In the following phenomenological study, we will simply choose
\beqa
g^\prime_{u_R} = - g^\prime_{d_R} \equiv g_q \,, \qquad 
g^\prime_{\chi_L} = g^\prime_{\chi_R} \equiv g_\chi \,.
\eeqa
Or, equivalently in terms of vector and axi-vector couplings, 
\beqa
g_u^V = \frac{1}{2} g_q \,, \qquad g_d^V = - \frac{1}{2} g_q \,,  \qquad 
g_u^A = \frac{1}{2} g_q \,, \qquad g_d^A = - \frac{1}{2} g_q \,,  \qquad
g_\chi^V = g_\chi \,, \qquad g_\chi^A = 0 \,.
\label{eq:couplings}
\eeqa

For the model at hand, the direct detection scattering cross section is dominated by the vector coupling to quarks and we have the vector coupling to protons $g_p^V = \frac{1}{2} g_q$ and to neutrons $g_n^V= -\frac{1}{2} g_q$, which is an iso-spin violating model. Neglecting the iso-spin form factor, we have the scattering cross section of dark matter off a nucleus $^A_Z N$ as
\beqa
\sigma^{\rm SI}_{\chi A} = \frac{(A-2 Z)^2}{\pi}\,\frac{g_q^2 \, g_\chi^2\,\mu^2_{\chi A} }{4\,M_{Z^\prime}^4} \,,
\eeqa
where $\mu_{\chi A}$ is the dark matter-nucleus reduced mass. The $(A-2Z)^2$ factor provides an additional suppression for experiments that have a target element with the same number of protons and neutrons. For CRESST-II~\cite{Angloher:2014myn}, among the three target elements both Oxygen and Calcium have suppressed scattering cross sections for the dominant isotope. The third element, Tungsten, only becomes sensitive when the dark matter mass is above 3 GeV~\cite{Angloher:2014myn}. Combined with the energy threshold of 0.6~keV, the direct detection constraints on the model are weak for $m_\chi \lesssim 1$~GeV, and we do not consider them any further.

The hadronic width of the $Z$ boson also constrains our model parameter space. Following a similar  calculation as in Refs.~\cite{Carone:1994aa,Carone:1995pu}, we have the summation of direct production of $Z \rightarrow \bar{q}q Z^\prime$ and the $Z\bar{q}q$ vertex correction to be
\beqa
\frac{\Delta \Gamma (Z\rightarrow \mbox{hadrons})}{\Gamma(Z \rightarrow q \bar{q})} = \frac{3}{16\pi^2} \, \frac{\sum_q [(g_Z^{V, q})^2 + (g_Z^{A, q})^2][(g_{Z^\prime}^{V, q})^2 + (g_{Z^\prime}^{A, q})^2] + 4  g_Z^{V, q} g_Z^{A, q} g_{Z^\prime}^{V, q} g_{Z^\prime}^{A, q} }{\sum_q [(g_Z^{V, q})^2 + (g_Z^{A, q})^2]} \,,
\eeqa
in the limit of $M_{Z^\prime} \ll M_Z$. Using the $Z$ hadronic decay branching ratio of $0.6991\pm0.0006$ from PDG~\cite{Agashe:2014kda}, we derive the constraint on the coupling, $g_q$, to be $g_q < 0.98$ at 90\% C.L. (shown in the green and shaded region of Fig.~\ref{fig:constrain-tevatron}).  There exist also contributions from the kinetic mixing of $Z^\prime$ with the hypercharge gauge boson, which depends on ultra-violet physics. For the kinetic mixing parameter of ${\cal O}(10^{-3})$, we have found a similar constraint as the one in Fig.~\ref{fig:constrain-tevatron}.

At colliders, the existing searches with the final state of mono-jet plus missing transverse energy can constrain the model parameter space in $g_q$ and $g_\chi$. For a light $Z^\prime$ around 1 GeV, it turns out that the Tevatron still provides the most stringent constraints~\cite{Aaltonen:2008hh,tevatron-monojet-link,Aaltonen:2012jb}. Using the analysis in Ref.~\cite{tevatron-monojet-link}, we recast the results to obtain constraints on the two couplings $g_q$ and $g_\chi$ in the left panel of Fig.~\ref{fig:constrain-tevatron}. Because the production cross sections from mono-$Z^\prime$ are subdominant compared to the mono-jet production, adding the mono-$Z^\prime$ signature does not change the constraints on couplings significantly.

At the LHC, the standard mono-jet searches with a cut on large missing transverse energy provide only worse limits than that from Tevatron. To search for light-$Z^\prime$ mediated dark matter production at the LHC, one needs to relax the missing transverse energy cut. Furthermore, one should also implement a different trigger to record more of the signal events. Just like the single hadronic $\tau$ trigger~\cite{Acosta:1955149}, one could define a similar mono-$Z^\prime$ jet trigger. In the right panel of Fig.~\ref{fig:constrain-tevatron}, we impose a cut of $p_T(Z^\prime) > 120$~GeV to search for dark matter together with a light $Z^\prime$ at the 14 TeV LHC. We also show the limits from applying jet-substructure cuts, assuming the default 50\% of the signal tag efficiency and 2\% background mistag efficiency. One can see that even though the mono-$Z^\prime$ signature has a small production cross section, the jet-substructure analysis can improve on the sensitivity beyond the mono-jet analysis.

\section{Discussion and Conclusions}
\label{sec:conclusion}
While all SM fermions have some charges under the SM gauge groups, the dark matter particle may be charged under its own gauge group. In this paper, we have studied the simplest case where the dark matter gauge group is an Abelian $U(1)^\prime$.  Within the realm of possible $U(1)^\prime$ models, we have concentrated on collider signatures of dark matter produced in association with a light $Z^\prime$ that mainly decays into a few hadrons.

For a heavier $Z^\prime$, the decay could be into two separate jets and the collider signature is $2j + E^{\rm miss}_T$ with a dijet resonance~\cite{Autran:2015mfa}. If the $Z^\prime$ boson can also decay into charged-leptons, a cleaner signature like $2\ell + E^{\rm miss}_T$ with a dilepton resonance should be searched for at the LHC experiments~\cite{Autran:2015mfa,Gupta:2015lfa}.  One could also extend the study here to a more complex model with a non-Abelian gauge group, which could be either spontaneously broken or confined in the infrared. A natural extension of our work could be a model-independent study of the collider signatures for the non-Abelian case. 

In this paper, we have considered two possibilities for how the dark matter and SM fermions are charged under the $Z^\prime$. For the secluded $Z^\prime$ model, where only the dark matter is charged, we have introduced effective operators to mediate dark matter-quark interactions. Those effective operators can easily be UV-completed by introducing another heavy gauge boson~\cite{An:2012va}. Before including the parton distribution function, the parton-level production cross section grows with the $p_T$ of the $Z^\prime$, so there is no issue with having a high enough trigger efficiency for signal sensitivity.

On the other hand, for our so-called ``public'' $Z^\prime$ model, increasing the $p_T$ cut on $Z^\prime$ does not improve the search sensitivity. The trigger at the LHC then becomes an issue if one wants to impose a looser cut on $p_T(Z^\prime)$. Fortunately, for a light $Z^\prime$, one could design a light $Z^\prime$ trigger to record the signal events. As shown in our jet sub-structure analysis, a light $Z^\prime$ jet is very similar to a hadronic tau, so the existing strategy for the hadronic tau trigger can be adapted to search for a class of dark matter models with a light $Z^\prime$ mediator. 

In summary, we have studied the collider signature of $U(1)^\prime$-charged dark matter models. Concentrating on a light GeV-scale $Z^\prime$ that mainly decays into a few hadrons, we have pointed out the new collider signature of a mono-$Z^\prime$ jet plus missing transverse energy at the LHC. We have performed a detailed jet-substructure analysis to demonstrate that tagging the $Z^\prime$ jet can dramatically reduce SM backgrounds and improve the limits on dark matter-quark interaction strengths. Both elastic and inelastic dark matter models have been studied in our paper and have similar results.  For the inelastic case, we find there is a better reach for lighter dark matter masses because of the enhanced two-body production when the $Z'$ is decay product of the heavier dark matter state. Comparing to the limits from the dark matter spin-dependent direct detection experiments, our proposed mono-$Z^\prime$ jet signature can provide a much more stringent constraint on the dark matter-nucleon interactions for a dark $U(1)^\prime$ gauge coupling order of unity, assuming that the operator that leads to dark matter production at colliders also dominates in dark matter-nucleon scattering. More generally, tagging on light $Z^\prime$ jets at the LHC can provide a new avenue to probe GeV-scale physics beyond the Standard Model.

\subsection*{Acknowledgments}
We would like to thank Bhawna Gomber, Andrew Larkoski, Matthew Low, Ran Lu, Iain Stewart, and Daniel Whiteson for useful discussion. YB and JB are supported by the U. S. Department of Energy under the contract DE-FG-02-95ER40896. This work was supported in part by the Kavli Institute for Cosmological Physics at the University of Chicago through grant NSF PHY-1125897 and an endowment from the Kavli Foundation and its founder Fred Kavli. We also thank the Aspen Center for Physics, under NSF Grant No. PHY-1066293, where the collaboration was formed.

\baselineskip14pt

\providecommand{\href}[2]{#2}\begingroup\raggedright\endgroup

 \end{document}